\documentclass[final,twocolumn,pre,aps,float,amsfonts,amssymb,showpacs,superscriptadress]{revtex4}
\usepackage{amsmath}
\usepackage{graphicx}
\usepackage{longtable}
\usepackage{epsfig}
\usepackage{subfigure}

\begin{document}

\title{Towards an experimental von K{\'a}rm{\'a}n dynamo: numerical studies 
for an optimized design}

\author{Florent Ravelet}
\affiliation{Service de Physique de l'{\'E}tat Condens{\'e}, DSM, CEA Saclay, CNRS URA 2464, 
91191 Gif-sur-Yvette, France}
\author{Jacques L{\'e}orat}
\email{Jacques.Leorat@obspm.fr}
\affiliation{LUTH, Observatoire de Paris-Meudon, 92195 Meudon, France}
\author{Arnaud Chiffaudel}
\email{arnaud.chiffaudel@cea.fr}
\affiliation{Service de Physique de l'{\'E}tat Condens{\'e}, DSM, CEA Saclay, CNRS URA 2464, 
91191 Gif-sur-Yvette, France}
\author{Fran\c{c}ois Daviaud}
\affiliation{Service de Physique de l'{\'E}tat Condens{\'e}, DSM, CEA Saclay, CNRS URA 2464, 
91191 Gif-sur-Yvette, France}

\date[To be published in Phys. Fluids: ]{\today}

\pacs{47.65+a, 91.25.Cw}

\begin{abstract} 

Numerical studies of a kinematic dynamo based on von K{\'a}rm{\'a}n type
flows between two counterrotating disks in a finite cylinder are
reported. The flow has been optimized using a water model experiment,
varying the driving impellers configuration. A solution leading to
dynamo action for the mean flow has been found. This solution may be
achieved in VKS2, the new sodium experiment to be performed in
Cadarache, France. The optimization process is described and
discussed, then the effects of adding a stationary conducting layer
around the flow on the threshold, on the shape of the neutral mode and
on the magnetic energy balance are studied. Finally, the possible
processes involved into kinematic dynamo action in a von K{\'a}rm{\'a}n
flow are reviewed and discussed. 
Among the possible processes we highlight the joint effect of the
boundary-layer radial velocity shear and of the Ohmic dissipation
localized at the flow/outer-shell boundary.

\end{abstract}

\maketitle

\section{Introduction} 

In an electrically conducting fluid, kinetic
energy can be converted into magnetic energy, if the flow is both of
adequate topology and sufficient strength. This problem is known as the
dynamo problem \cite{moffat78}, and is a magnetic
seed-field instability. The equation describing the behavior of the magnetic
induction field $\bf B$ in a fluid of resistivity $\eta$ under the
action of a velocity field $\bf v$ writes in a dimensionless form:

\begin{equation}
\frac{\partial \bf B}{\partial t}=\nabla \times (\bf v \times \bf B) +
\frac{\eta}{\cal V^* \cal L^*} \nabla^2 \bf B  \label{eq:kde}
\end{equation}

where $\cal L^*$ is a typical length scale and $\cal V^*$ a typical
velocity scale. In addition, one must take into account the divergence-free of $\bf B$, the electromagnetic boundary
conditions and the Navier-Stokes equations governing
the fluid motion, including the back-reaction of the magnetic
field on the flow through the Lorentz force. 

The magnetic Reynolds number $R_m={\cal V^* \cal L^*}{\eta}^{-1}$, which
compares the advection to the Ohmic diffusion, controls the instability.
Although this problem is simple to set, it is still open. While some
flows lead to the dynamo instability with a certain threshold $R_m^c$,
other flows do not, and anti-dynamo theorems are not sufficient to
explain this sensitivity to flow geometry \cite{moffat78}. The two
recent experimental success of Karlsruhe and Riga
\cite{gailitis00,Stieglitz01,gailitis01,gailitis02,muller04} are in
good agreement with analytical and numerical calculations
\cite{stefani99,tilgner02,tilgnerbusse02,plunian02}; these two dynamos
belong to the category of constrained dynamos: the flow is forced in
pipes and the level of turbulence remains low. However, the saturation
mechanisms of a dynamo are not well known, and the role of turbulence on
this instability remains misunderstood
\cite{gailitis03,cattaneo96,sweet01,petrelis01,ponty04,thesepetrelis,
fauve03}. 

The next generation of experimental homogeneous unconstrained dynamos
(still in progress, see for example Frick {\it et al.}, Shew {\it et
al.}, Mari{\'e} {\it et al.} and O'Connell {\it et al.} in the
Carg{\`e}se 2000 workshop proceedings \cite{cargese}) might provide
answers to these questions. The VKS liquid-sodium experiment held in
Cadarache, France \cite{marie02,bourgoin02,petrelis02} belong to this
category. The VKS experiment is based on a class of flows called von
K{\'a}rm{\'a}n type flows. In a closed cylinder, the fluid is inertially
set into motion by two coaxial counterrotating impellers fitted with
blades. This paper being devoted to the hydrodynamical and
magnetohydrodynamical properties of the mean flow, let us first describe
briefly the phenomenology of such mean flow. Each impeller acts as a
centrifugal pump: the fluid rotates with the impeller and is expelled
radially by centrifugal effect. To ensure mass conservation the fluid is
pumped in the center of the impeller and recirculates near the cylinder
wall. In the exact counterrotating regime, the mean flow is divided into
two toric cells separated by an azimuthal shear layer. Such a mean flow
has the following features, known to favor dynamo action: differential
rotation, lack of mirror symmetry and presence of a hyperbolic
stagnation point in the center of the volume. In the VKS experimental
devices, the flow, inertially driven at kinetic Reynolds number up to
$10^7$ (see below), is highly turbulent. As far as full numerical MHD
treatment of realistic inertially driven high-Reynolds-number flows
cannot be carried out, this study is restricted to the kinematic dynamo
capability of von K\'arm\'an mean flows.

Several measurements of induced fields have been performed in the first
VKS device (VKS1) \cite{bourgoin02}, in rather good agreement with
previous numerical studies \cite{marie03}, but no dynamo was seen: in
fact the achievable magnetic Reynolds number in the VKS1 experiment
remained below the threshold calculated by Mari{\'e} {\em et al.}
\cite{marie03}. A larger device ---VKS2, diameter $0.6$~m and $300$~kW
power supply--- is under construction. 
The main generic properties of mean flow dynamo action have been
highlighted by Mari{\'e} {\em et al.} \cite{marie03} on two different
experimental von K\'arm\'an velocity fields. Furthermore, various
numerical studies in comparable spherical flows confirmed the strong effect of
flow topology on dynamo action \cite{dudley89,forest02}. In the
experimental approach, lots of parameters can be varied, such as the
impellers blade design, in order to modify the flow features.
In addition, following Bullard \& Gubbins \cite{bullard77}, several
studies suggest to add a layer of stationary conductor around the flow
to help the dynamo action. All these considerations lead us to consider
the implementation of a static conducting layer in the VKS2 device and
to perform a careful optimization of the mean velocity field by a
kinematic approach of the dynamo problem. 

Looking further towards the real VKS2 experiment, one should discuss the major remaining physical unexplored feature: the role of hydrodynamical turbulence. Such turbulence in an inertially-driven closed flow will be very far from homogeneity and isotropy. The presence of hydrodynamical small scale turbulence could act in two different ways: on the one hand, it may increase the effective magnetic diffusivity, inhibiting the dynamo action \cite{reighard01}. On the other hand, it could help the dynamo through small-scale $\alpha$-effect \cite{krause80}. Moreover, the presence of a turbulent mixing layer between the two counterrotating cells may move the instantaneous velocity field away from the time-averaged velocity field for large time-scales \cite{marie04pof}. As the VKS2 experiment is designed to operate above the predicted kinematic threshold presented in this paper, it is expected to give an experimental answer to the question about the role of turbulence on the instability. Furthermore, if it exhibits dynamo, it will allow to study the dynamical saturation regime which is outside the present paper scope.

In this article, we report the optimization of the time-averaged flow in
a von K\'arm\'an liquid sodium experiment. We design a solution which can
be experimentally achieved in VKS2, the new device held in Cadarache,
France. This solution particularly relies on the addition of a static conducting
layer surrounding the flow. The paper is organized as follows. In Section~\ref{sec:tools} we first
present the experimental and numerical techniques that have been used.
In Section~\ref{sec:optim}, we present an overview of the
optimization process which lead to the experimental configuration chosen
for the VKS2 device. We study the influence of the shape of the
impellers both on the hydrodynamical flow properties and on the onset of
kinematic dynamo action. 
In Section~\ref{sec:layer}, we focus on the understanding
of the observed kinematic dynamo on a magnetohydrodynamical point of view: we
examine the structure of the eigenmode and the effects of an outer
conducting boundary. 
Finally, in Section~\ref{sec:conjec}, we review some possible
mechanisms leading to kinematic dynamo action in a von K{\'a}rm{\'a}n
flow and propose some conjectural explanations based on our
observations.

\section{Experimental and numerical tools}
\label{sec:tools}

\subsection{What can be done numerically}

The bearing of numerical simulations in the design of experimental fluid
dynamos deserves some general comments. Kinetic Reynolds numbers of such
liquid sodium flows are typically $10^7$, well beyond any conceivable
direct numerical simulation. Moreover, to describe effective MHD
features, it would be necessary to manage very small magnetic Prandtl
numbers, close to $10^{-5}$, a value presently out of computational
feasibility. Several groups are progressing in this way on model flows,
for example with Large Eddy Simulations \cite{ponty04} which can reach
magnetic Prandtl numbers as low as $10^{-2}$-- $10^{-3}$.
Another strong difficulty arises from the search of realistic magnetic boundary
conditions treatment which prove in practice also to be difficult to
implement, except for the spherical geometry.

An alternative numerical approach is to introduce a given flow in the magnetic
induction equation \ref{eq:kde} and to perform \lq\lq kinematic dynamo\rq\rq
computations. This flow can be either analytical \cite{dudley89,tilgner02}, computed by pure
hydrodynamical simulations (which may now be performed with Reynolds
numbers up to a few thousands), or measured in laboratory water models
\cite{forest02,marie03} by Laser Doppler velocimetry (LDV) or by
Particle Imaging Velocimetry (PIV).
Such measurements lead to a map of the time-averaged flow and to main
properties of the fluctuating components: turbulence level,
correlation times, etc... 
Kinematic dynamo computations have been successfully used to describe or
optimize the Riga \cite{stefani99} and Karlsruhe \cite{tilgner02}
dynamo experiments.

We will follow here the kinematic approach using the time-averaged flow
measured in a water model at realistic kinetic Reynolds number.
Indeed, potentially important features such as velocity fluctuations will not be
considered. Another strong limitation of our pragmatic kinematic
approach is its linearity: computations may predict if an initial seed
field grows, but the study of the saturation regime will rely
exclusively on the results of the real MHD experiment VKS. 

\subsection{Experimental measurements}

In order to measure the time-averaged velocity field ---hereafter simply
denoted mean field--- we use a water-model experiment which is a
half-scale model of the VKS2 sodium device. The experimental setup,
measurement techniques, and methods are presented in detail in Refs.
\cite{marie03,theselouis}. However, we present below an overview of our
experimental issues and highlight the evolutions with respect to those
previous works.

We use water as working fluid for our study, noting that its
hydrodynamical properties at $50^oC$ (kinematic viscosity $\nu$ and
density $\rho$) are very close to sodium properties at $120^oC$. 

\begin{figure}[tbp]
\begin{center}
\includegraphics[clip,width=7cm]{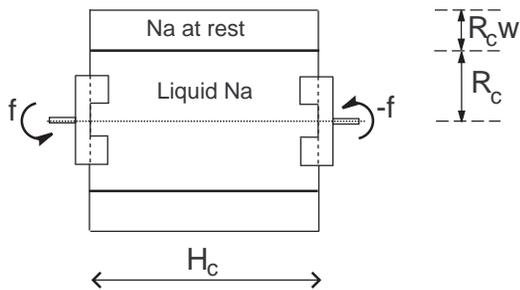}%
\end{center} 
\caption{Sketch of the VKS2 experiment. The container radius $R_c$ is taken as unit scale. $w$ is the dimensionless thickness of sodium at rest.}
\label{fig:sketch} 
\end{figure}

A sketch of the von K\'arm\'an experiments is presented in
Fig.~\ref{fig:sketch}. The cylinder is of radius $R_c$ and height $H_c =
1.8 R_c$. In the following, all the spatial quantities are given in
units of $R_c = {\cal L^*}$. The hydrodynamical time scale is based on
the impeller driving frequency $f$: if ${{{\bf V}}}$ is the measured
velocity field for a driving frequency $f$, the dimensionless mean
velocity field is thus ${\bf {v}} = (2 \pi R_cf)^{-1} {\bf V}$.

The integral kinetic Reynolds number $Re$ is typically $10^6$ in the
water-model, and $10^7$ in the sodium device VKS2. The inertially driven
flow is highly turbulent, with velocity fluctuations up to $40$ percent
of the maximum velocity \cite{bourgoin02, marie03}. In the water model, we measure
the time-averaged velocity field by Laser Doppler Velocimetry (LDV).
Data are averaged over typically $300$ disk rotation periods. We have
performed measurements of velocity in several points for several driving
frequencies: as expected for so highly turbulent a
flow, the dimensionless velocity $\bf{v}$ does not depend on the
integral Reynolds number $Re={\cal V^* \cal L^*}\nu^{-1}$
\cite{frisch95}.

Velocity modulations at the blade frequency have been observed only in
and very close to the inter-blade domains. These modulations are thus
time-averaged and we can consider the mean flow as a solenoidal
axisymmetric vector field \cite{ravelet02}. So the toroidal part of the velocity field
$V_\theta$ (in cylindrical coordinates) and the poloidal part $(V_z,
V_r)$ are independent. 

\begin{figure}[htb!]
\begin{center}
\includegraphics[clip,width=8cm]{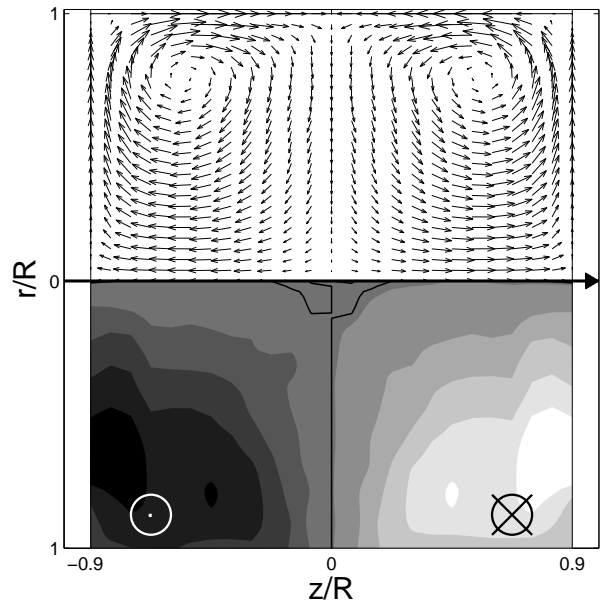}%
\end{center} 
\caption{Dimensionless mean velocity field measured by LDV and symmetrized for kinematic dynamo simulations. Cylinder axis is horizontal. Arrows correspond to poloidal part of the flow, color code to toroidal part. We use
cylindrical coordinates $(r,\theta,z)$, with origin at the center of the cylinder.}
\label{fig:ldv1} 
\end{figure}

In the water model experiment dedicated to the study reported in this
paper, special care has been given to the measurements of velocity
fields, especially near the blades and at the cylinder wall, where the
measurement grid has been refined. The mechanical quality of the
experimental setup ensures good symmetry of the mean velocity fields
with respect to rotation of $\pi$ around any diameter passing through
the center of the cylinder ($\cal R_{\pi}$-symmetry). The fields
presented in this paper are thus symmetrized by $\cal R_{\pi}$ with no
noticeable changes in the profiles but with a slightly improved spatial
signal-to-noise ratio. With respect to Ref.~\cite{marie03}, the velocity
fields are neither smoothed, nor stretched to different aspect ratios. 

Fig. \ref{fig:ldv1} shows the mean flow produced by the optimal
impeller. The mean flow
respects the phenomenology given in the Introduction: it is made of
two toroidal cells separated by a shear layer, and two poloidal
recirculation cells. High velocities are measured
in the whole volume: the inertial stirring is actually very efficient.
Typically, the average over the flow volume of the mean velocity field
is of order of $0.3 \times (2 \pi R_c f)$.

In addition to velocity measurements, we perform global power
consumption measurements: torques are measured through the current
consumption in the motors given by the servo drives and have been
calibrated by calorimetry.

\subsection{Kinematic dynamo simulations}

Once we know the time-averaged velocity field, we integrate the
induction equation using an axially periodic kinematic dynamo code,
written by J.~L{\'e}orat \cite{leorat94}. The code is pseudo-spectral in
the axial and azimuthal directions whether radial dependence is treated
by high-order finite difference scheme. The numerical resolution
corresponds to a grid of $48$ points in the axial direction, $4$ points
in the azimuthal direction (corresponding to wave numbers $m=0,\pm1$)
and $51$ points in the radial direction for the flow domain. This
spatial grid is the common basis of our simulations and has been refined
in some cases. The time scheme is second-order Adams-Bashforth with
diffusive time unit $t_d=R_c^2 \eta^{-1}$. Typical time step is
$5.10^{-6}$ and simulations are generally carried out over $1$ time
unit. 

Electrical conductivity and magnetic permeability are homogeneous and
the external medium is insulating. Implementation of the magnetic
boundary conditions for a finite cylinder is difficult, due to the
non-local character of the continuity conditions at the boundary of the
conducting fluid. On the contrary, axially periodic boundary conditions
write down easily, since the harmonic external field has then an
analytical expression. We choose thus to look for axially periodic
solutions, using a relatively fast code, which allows to perform
parametric studies. To validate our choice, we compared our results with
results from a finite cylinder code (F.~Stefani, private communication)
for some model flows and a few experimental flows. It happens that in
all these cases, the periodic and the finite cylinder computations give
comparable results. This remarkable agreement may be due to the peculiar
flow and to the magnetic eigenmodes symmetries: we do not claim that it
may be generalized to other flow geometries. Indeed, the numerical
elementary box consists of two mirror-symmetric experimental velocity
fields in order to avoid strong velocity discontinuities along $z$ axis.
The magnetic eigenmode could be either symmetric or antisymmetric
towards this artificial mirror symmetry \cite{knobloch96}. In the quasi
totality of our simulations, the magnetic field is mirror-antisymmetric,
and we verify that no axial currents cross the mirror boundary. The few
exotic symmetric cases we encountered cannot be used for experiments
optimization.

Further details on the code can be found in Ref.~\cite{leorat94}. We use
mirror-antisymmetric initial magnetic seed field optimized for fast
transient~\cite{marie03}. Finally, we can act on the electromagnetic boundary
conditions by adding a layer of stationary conductor of dimensionless
thickness $w$, surrounding the flow exactly as in the experiment
(Fig.~\ref{fig:sketch}). This extension is made keeping the grid radial
resolution constant ($51$ points in the flow region).

\section{Optimization of the VKS experiment}
\label{sec:optim}

\subsection{Optimization process}

The goal of our optimization process is to find the impeller whose mean
velocity field leads to the lowest $R_m^c$ for the lowest power cost. 
We have to find a solution feasible in
VKS2, {\em i.e.} with liquid sodium in a $0.6$~m diameter cylinder with
$300$~kW power supply. We
performed an iterative optimization loop: for a given configuration, we
measure the mean velocity field and the power consumption. Then we
simulate the kinematic dynamo problem. We try to identify features
favoring dynamo action and modify parameters in order to reduce the
threshold and the power consumption and go back to the loop. 

\subsection{Impeller tunable parameters.}

\begin{figure}[htbp!]
\begin{center}
\includegraphics[clip,width=4cm]{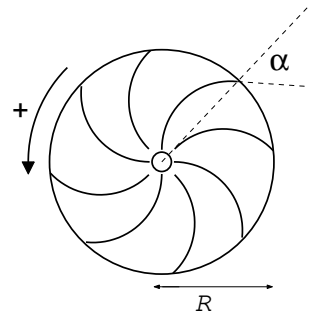}%
\end{center} 
\caption{Sketch of the impeller parameters. $R$ is the dimensionless radius, $\alpha$ the blade curvature angle. The sign of $\alpha$ is determined by the sense of rotation: positive when rotated anticlockwise.}
\label{fig:turbine} 
\end{figure}

The impellers are flat disks of radius $R$ fitted with $8$
blades of height $h$. The blades are arcs of circle with a curvature radius $C$
and are radial at the center of the disks. We use the angle
$\alpha=\arcsin(\frac{R}{2C})$ to label the different curvatures
(see Fig.~\ref{fig:turbine}). For straight blades $\alpha=0$. By convention, we use positive values to label
the direction corresponding to the case where the fluid is set into motion by the convex
face of the blades. In order to study the opposite curvature
($\alpha<0$) we just rotate the impeller in the other direction. The two
counterrotating impellers are separated by $H_c$, the height of the
cylinder. We fixed the aspect ratio ${H_c}/{R_c}$ of the flow volume to
$1.8$ as in VKS device. In practice we successively examine the effects of each
parameter $h$, $R$ and $\alpha$ on some global quantities
characterizing the mean flow. We then varied the parameters one by
one, until we found a relative optimum for the dynamo threshold. We tested $12$
different impellers, named TMxx, with three radii ($R=0.5, 0.75 \; \& \; 0.925$), various curvature angles $\alpha$ and different
blade heights $h$.

\subsection{Global quantities and scaling relations} 

We know from empirical results
\cite{marie03,dudley89,forest02} that the poloidal to toroidal ratio $\Gamma$ of
the flow has a great impact on the dynamo threshold. Moreover, a purely
toroidal flow is unable of sustained dynamo action \cite{bullard54,
backus58}, while it is possible for a purely poloidal flow
\cite{love96,proctor04}. We also notice that, for a Ponomarenko flow,
the pitch parameter plays a major role
\cite{stefani99,thesepetrelis,fauve03}. All these results lead us to
first focus on the ratio $$\Gamma=\frac{\langle P \rangle}{\langle T
\rangle}$$ where $\langle P \rangle$ is the spatially averaged value of
the poloidal part of the mean flow, and $\langle T \rangle$ the average
of the toroidal part. 

Another quantity of interest is the velocity factor $\cal{V}$: the
dimensionless maximum value of the velocity. In our simulations, the
magnetic Reynolds number $R_m$ is based on the velocity factor, {\em
i.e.} on a typical {\em measured} velocity in order to take into account the stirring efficiency: $${\cal{V}} \; = \;
\frac{{\rm{max}}(||{\bf{{V}}}||)}{ 2 \; \pi \; {\rm{R_c \; f}}}$$ $$R_m
\; = \; 2 \; \pi \; {\rm{R_c^2 \; f}} \; \cal{V} \; / \; \eta$$

We also define a power coefficient $K_p$ by dimensional analysis. We
write the power $\mathcal P$ given by a motor to sustain the flow as
follows: $${\mathcal P} \; = \; K_p(Re,~geometry) \rho \; R_c^5 \;
\Omega^3$$ with $\rho$ the density of the fluid and $\Omega\; = \; 2 \pi
f$ the driving pulsation. We have checked \cite{theselouis} that $K_p$
does not depend on the Reynolds number $Re$ as expected for so highly
turbulent inertially driven flows \cite{frisch95}.

The velocity factor measures the stirring efficiency: the greater
${\cal{V}}$, the lesser the rotation frequency needed to reach a given velocity.
Besides, the lesser $K_p$, the lesser the power to sustain a given
driving frequency. The dimensionless number which we need to focus on
compares the velocity effectively reached in the flow to the power
consumption. We call it the ${\rm MaDo}$ number: $$\rm{MaDo} \; = \;
\frac{\cal{V}}{K_p^{1/3}}$$ The greater ${\rm MaDo}$, the lesser the
power needed to reach a given velocity ({\it i.e.} a given magnetic
Reynolds number). The ${\rm MaDo}$ number is thus a hydrodynamical
efficiency coefficient. To make the VKS experiment feasible at
laboratory scale, it is necessary both to have great ${\rm MaDo}$ numbers and low critical magnetic Reynolds numbers 
$R_m^c$. The question laying under the process of optimization is to know if we could on
the one hand find a class of impellers with mean flows exhibiting dynamo
action, and on the other hand if we could increase the ratio ${\rm
MaDo}/{R_m^c}$. It means that we have to look both at the global
hydrodynamical quantities and at the magnetic
induction stability when varying the impellers tunable parameters $h$, $R$
and~$\alpha$. 

\begin{figure}[tbp]
\begin{center}
\includegraphics[clip,width=7.5cm]{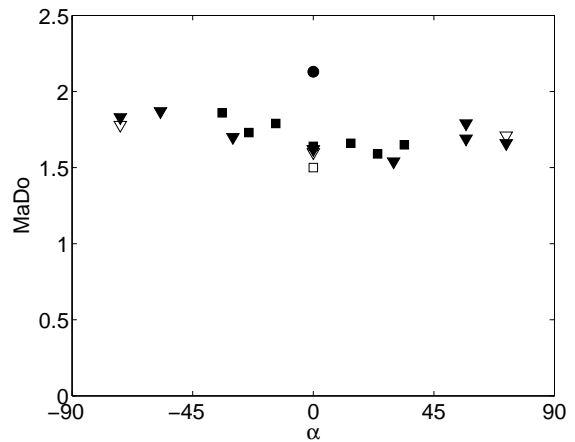}%
\end{center} 
\caption{$\rm MaDo$ number vs $\alpha$ for all the impellers we have tested. $R= 0.925 (\blacktriangledown)$, $R= 0.75 (\blacksquare)$ and $R= 0.5 (\bullet)$. Closed symbols: $h=0.2$. Open symbols: $h \leq 0.1$}
\label{fig:mado} 
\end{figure}

Fig. \ref{fig:mado} presents $\rm{MaDo}$ for the whole set of impellers.
For our class of impellers, the $\rm MaDo$ number remains of the same
order of magnitude within $\pm 10\%$. Only the smallest diameter
impeller ($R=0.5$) exhibits a slightly higher value. In the ideal case
of homogeneous isotropic turbulence, far from boundaries, we can show
that what we call the $\rm{MaDo}$ number is related to the Kolmogorov
constant $C_K \simeq 1.5$ \cite{lesieur}. The Kolmogorov constant is
related to the kinetic energy spatial spectrum:

$$E(k)=C_K\;\epsilon^{2/3}\;k^{-5/3}$$

\noindent where $\epsilon$ is the massic dissipated power, and $k$ the wave
number. If we assume that $\epsilon$ is homogeneous ---${\cal P}$ being
the total dissipated power we measure--- we have: $$\epsilon=\frac{\cal
P}{\rho \pi R_c^2H_c}$$ Using the definition $$\frac{1}{2}\langle v^2
\rangle=\int E(k)dk$$ and assuming $\frac{1}{2}\langle v^2 \rangle
\simeq \frac{1}{2}{\cal V}^2$ and using the steepness of the spectrum, we
obtain: $$E(k_0)=\frac{1}{3} {\cal V}^2 k_0^{-1}$$ with $k_0=2\pi/R_c$
the injection scale. Then the relation between the $\rm{MaDo}$ number
and $C_K$ would be: $$\rm{MaDo}^2 \simeq 3\pi^{-4/3}
(\frac{H_c}{R_c})^{-2/3} C_K  \simeq 0.44 C_K$$ {\it i.e.}, with $C_K=1.5$,
we should have, for homogeneous isotropic turbulence
$\rm{MaDo}\simeq0.81$. In our closed system with blades, we recover the
same order of magnitude, and the fact that $\rm{MaDo}$ does merely not depend on the
driving system. Thus, there is no obvious optimum for the hydrodynamical
efficiency. Between various impellers producing dynamo action, the choice will
be dominated by the value of the threshold $R_m^c$.

Let us first get rid of the effect of the blade height $h$. The power
factor $K_p$ varies quasi-linearly with $h$. As $\rm MaDo$ is almost
constant, smaller $h$ impellers require higher rotation frequencies, rising
technical difficulties. We choose $h=0.2$, a compromise between stirring
efficiency and the necessity to keep the free volume sufficiently large.

\subsection{Influence of the poloidal/toroidal ratio $\Gamma$}

In our cylindrical von K\'arm\'an flow without conducting layer ($w=0$), there seems to be an optimal value
for $\Gamma$ close to $0.7$.
Since the mean flow is axisymmetric and divergence-free, the ratio
$\Gamma$ can be changed numerically by introducing an arbitrary
multiplicative factor on, say, the toroidal part of the velocity field.
In the following, $\Gamma_0$ stands for the experimental ratio for the
measured mean velocity field, whereas $\Gamma$ stands for the numerically
adjusted ratio. 

\begin{figure}[htbp!]
\begin{center}
\includegraphics[clip,width=8cm]{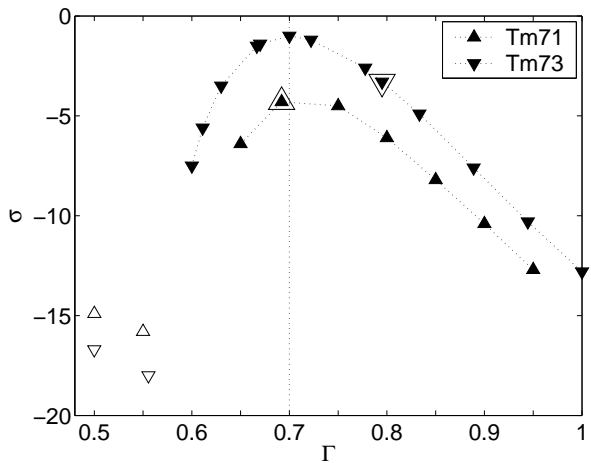}%
\end{center} 
\caption{Magnetic energy growth rate $\sigma$ vs. numerical ratio~$\Gamma$. $R_m=100$, $w=0$. Simulations performed for two different mean velocity fields (impellers TM71 ($\blacktriangle$) and TM73 ($\blacktriangledown$) of radius $R=0.75$). Larger symbols correspond to natural $\Gamma_0$ of the impeller. Vertical dashed line corresponds to optimal $\Gamma=0.7$. Closed symbols stand for stationary regimes, whereas open symbols stand for oscillating regimes for $\Gamma \lesssim 0.6$.}
\label{fig:gammanumw0} 
\end{figure}

In Fig. \ref{fig:gammanumw0}, we plot the magnetic energy growth rate
$\sigma$ (twice the magnetic field growth rate) for different values of
$\Gamma$, for magnetic Reynolds number $R_m=100$ and without conducting
layer ($w=0$). The two curves correspond to two different mean velocity
fields which have been experimentally measured in the water model (they correspond to the 
TM71 and TM73 impellers, see table \ref{tab:zob} for their
characteristics). We notice that the curves show the same bell shape
with maximum growth rate at $\Gamma \simeq 0.7$, which confirms
the results of Ref.~\cite{marie03}. 

\begin{figure}[htp!]
\begin{center}
\includegraphics[clip,width=8cm]{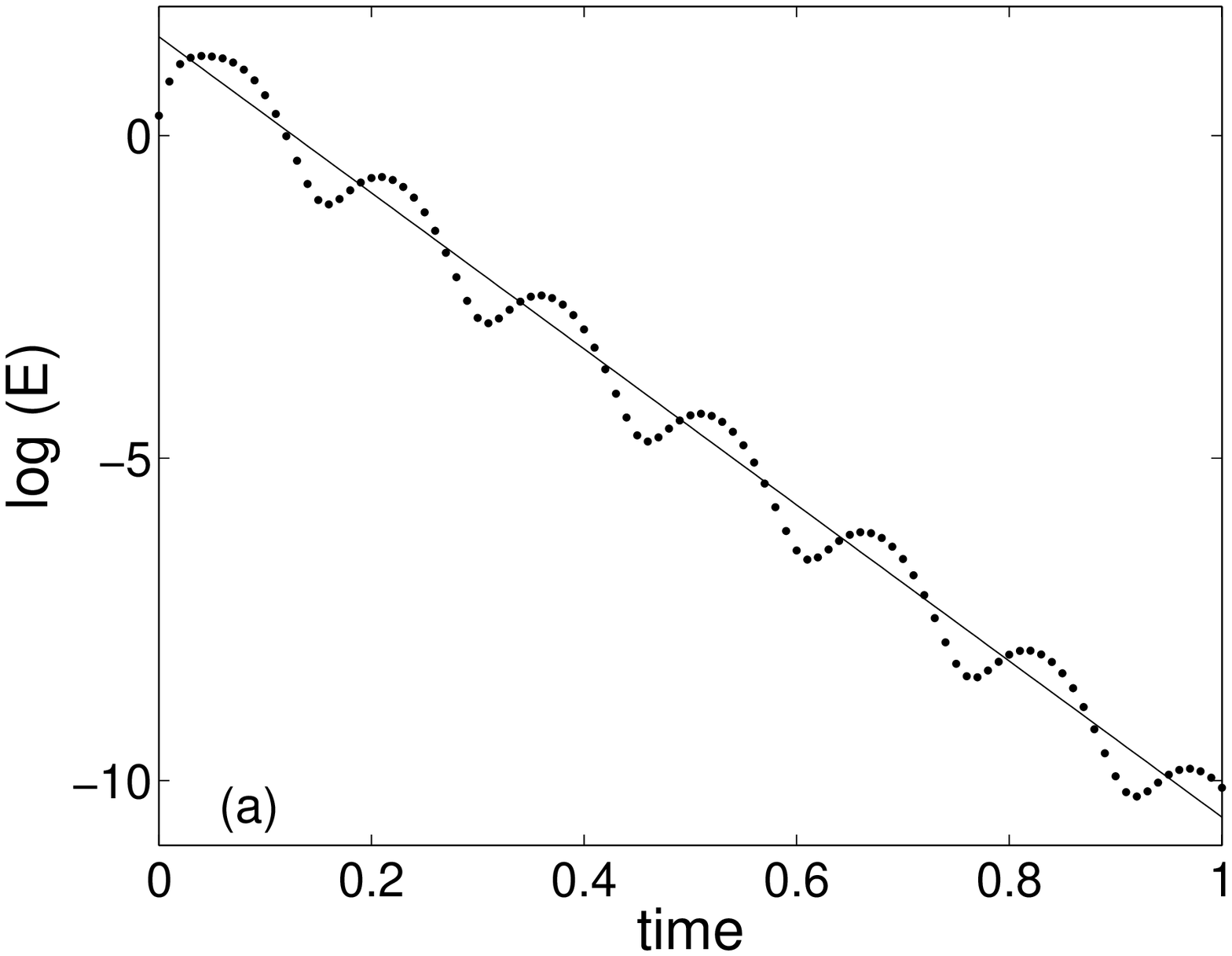}

\includegraphics[clip,width=8cm]{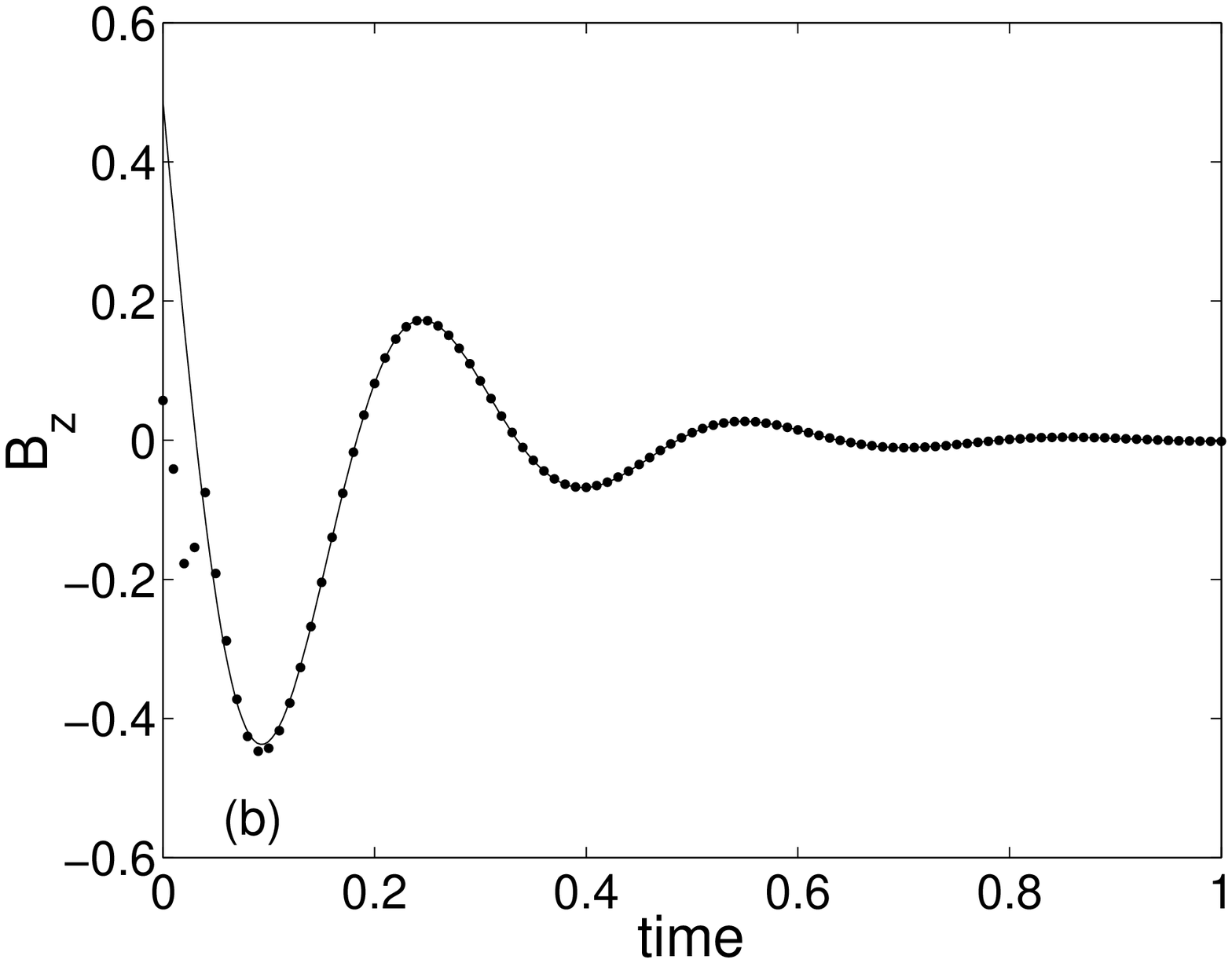}
\end{center} 
\caption{Typical damped oscillating regime for impeller TM70 at $\Gamma=0.5$, $w=0$, $R_m=140$. (a): temporal evolution of the magnetic energy $E=\int \bf B^2$. Straight line is a linear fit of the form $E(t)=E_0 \; exp(\sigma t)$ and gives the temporal growth rate $\sigma=-12.1$. (b): temporal evolution of the $z$ component of $\bf B$ at the point $r=0.4, \theta=0, z=-0.23$ with a nonlinear fit of the form: $B_z(t)=a \; exp (\sigma t/2) \; cos(\omega t+\phi)$ which gives $\sigma=-12.2$ and $\omega=20.7$.}
\label{fig:energietempo} 
\end{figure}

For $\Gamma \lesssim 0.6$, oscillating damped regimes (open symbols in
Fig. \ref{fig:gammanumw0}) are observed. We plot the temporal evolution
of the magnetic energy in corresponding case in Fig.
\ref{fig:energietempo}: these regimes are qualitatively different from
the oscillating regimes already found in \cite{marie03} for {\em non}
$\cal R_{\pi}$-symmetric $\Gamma=0.7$ velocity fields, consisting of one
mode with a complex growth rate: the magnetic field is a single
traveling wave, and the magnetic energy, integrated over the volume,
evolves monotonically in time.

In our case, the velocity field is $\cal R_{\pi}$-symmetric. It is known
that symmetries strongly constrain the nature of eigenvalues and eigenmodes
of linear stability problems. Due to $\cal R_{\pi}$-symmetry invariance
of the evolution operator for the magnetic field, two type of solutions
are allowed \cite{knobloch96}: 
\begin{itemize} 
\item One $\cal R_{\pi}$-symmetric eigenmode with a real eigenvalue. The corresponding
bifurcation is steady. 
\item Two eigenmodes images one of the other by
$\cal R_{\pi}$, associated with complex-conjugate eigenvalues.
\end{itemize} 
For $\Gamma \agt 0.6$, we always observed stationary
regimes. Otherwise, for $\Gamma \lesssim 0.6$, starting the temporal
integration with an initial condition for the magnetic field which has
non vanishing projection on both eigenmodes, we obtain a mix of two
modes with complex-conjugate growth rates and the magnetic energy decays
exponentially while pulsating (Fig. \ref{fig:energietempo}). The same
feature has been reported for analytical ``$s_2^0t_2^0-like$ flows'' in
a cylindrical geometry with a Galerkin analysis of neutral modes and
eigenvalues for the induction equation \cite{marie04}. A major interest
of the latter method is that it gives the structure of the modes: one mode is
localized near one impeller and rotates with it, the other being
localized and rotating with the other impeller. Growing oscillating
dynamos are rare in our system: a single case has been observed, for
TM71$(-)$ ($\Gamma_0=0.53$) with a $w=0.4$ conducting layer at $R_m=215$
($R_m^c=197$, see table \ref{tab:zob}). Such high a value for the
magnetic Reynolds number is out of the scope of our experimental study,
and is close to the practical upper limit of the numerical code. 

Experimental dynamo action will thus be searched in the stationary
regimes domain $\Gamma \agt 0.6$. Without conducting layer, we have to
look for the optimal impeller around $\Gamma_0 \simeq 0.7$.


\subsection{Effects of the impeller radius $R$}

\begin{figure}[ht!]
\begin{center}
\includegraphics[clip,width=8cm]{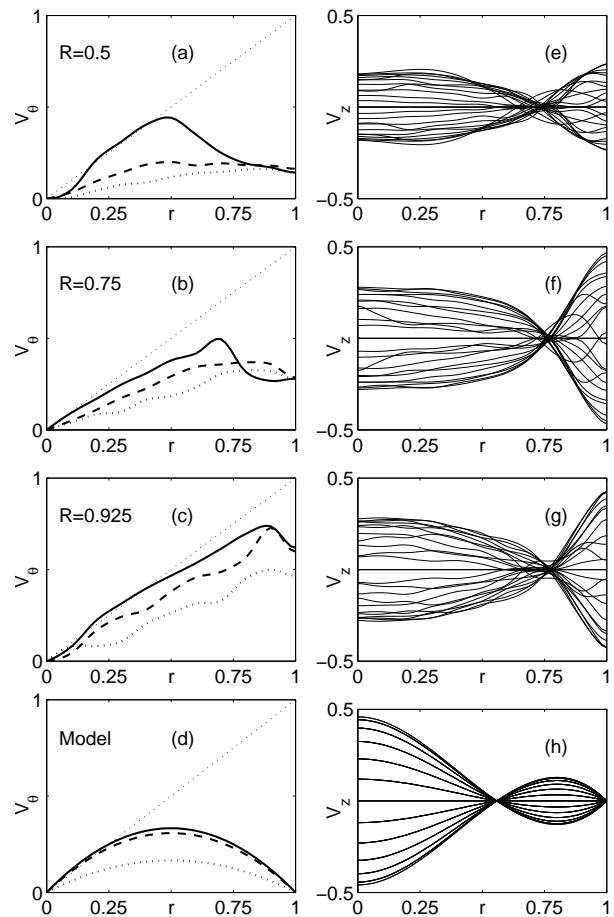}%
\end{center} 
\caption{Radial profiles of toroidal velocity $v_\theta$ ((a)--(d)) for $z=0.3$ (dotted line),  $0.675$ (dashed line),  $\& \; 0.9$ (solid line); and axial velocity $v_z$ ((e)--(h)) for various equidistant $z$ between the two rotating disks. From top to bottom: experimental flow for (a-e): $R=0.5$, (b-f): $R=0.75$, (c-g): $R=0.925$ impeller and (d-h): model analytical flow given by equations (1) (see discussion below p. \pageref{eq:mnd}).}
\label{fig:radialV} 
\end{figure}

One could {\it a priori} expect that a very large impeller is favorable
to the hydrodynamical efficiency. This is not the case. For impellers with
straight blades, $\rm MaDo$ slightly decreases with $R$: for
respectively $R=0.5$, $0.75$ and $0.925$, we respectively get
$\rm MaDo=2.13$, $1.64$ and $1.62$. This tendency is below the
experimental error. We thus consider that $\rm{MaDo}$ does not depend on
the impeller.

Nevertheless one should not forget that $\cal V$ varies quasi-linearly
with impeller radius $R$: if the impeller becomes smaller it must rotate faster to
achieve a given value for the magnetic Reynolds number, which may again
cause mechanical difficulties. We do not explore radii $R$ smaller than
$0.5$.

Concerning the topology of the mean flow, there is no noticeable effects
of the radius $R$ on the poloidal part. We always have two toric
cells of recirculation, centered at a radius $r_p$ close to $0.75 \pm
0.02 $ and almost constant for all impellers (see right part of Fig.
\ref{fig:radialV}). The fluid is pumped to the impellers for $0<r<r_p$
and is reinjected in the volume $r_p<r<1$. This can be interpreted as a
geometrical constraint to ensure mass conservation: the circle of radius
$r=\frac{\sqrt{2}}{2}$~(very close to $ 0.75$) separates the unit disk
into two regions of same area.
 
The topology of the toroidal part of the mean flow now depends on the
radius of the impeller. The radial profile of $v_{\theta}$ shows
stronger departure from solid-body rotation for smaller $R$ (left part
of Fig.~\ref{fig:radialV}): this will be emphasized in the discussion.
We performed simulations for three straight blades impellers of radii
$R=0.5$, $R=0.75$ and $R=0.925$; without conducting shell ($w=0$) and with a conducting layer of thickness $w=0.4$. We have
integrated the induction equation for the three velocity fields
numerically set to various $\Gamma$ and compared the growth rates. The
impeller of radius $R=0.75$ close to the radius of the center of the
poloidal recirculation cells systematically gets greatest growth rate.
So, this radius $R=0.75$ has been chosen for further investigations.

\subsection{Seek for the optimal blade curvature}

\begin{figure}[hbp!]
\begin{center}
\includegraphics[clip,width=8cm]{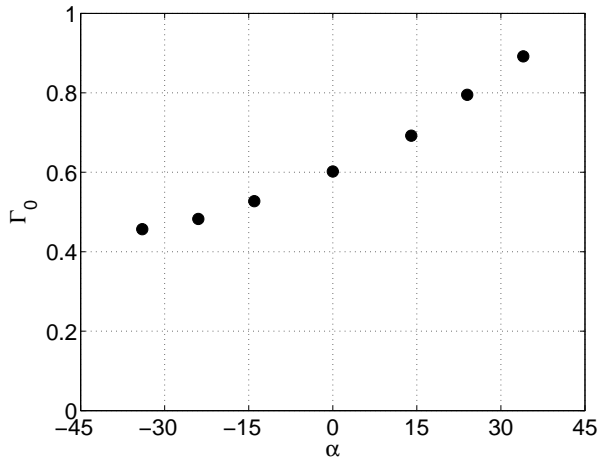}%
\end{center} 
\caption{$\Gamma_0$ vs $\alpha$ for four impellers of radius $R=0.75$ rotated in positive and negative direction (see Table \ref{tab:zob}).}
\label{fig:gammaphi75} 
\end{figure}

The hydrodynamical characteristics of the impellers of radius $R=0.75$ are given in table \ref{tab:zob}. For increasing blade curvature
the average value of the poloidal velocity $\langle P \rangle$ increases
while the average value of the toroidal velocity $\langle T \rangle$
decreases: the ratio $\Gamma_0$ is a continuous growing function of
curvature $\alpha$ (Fig.~\ref{fig:gammaphi75}). A phenomenological
explanation for $\langle T \rangle$ variation can be given. The fluid
pumped by the impeller is centrifugally expelled and is constrained to
follow the blades. So, it exits the impeller with a velocity almost
tangent to the blade exit angle $\alpha$. Thus, for $\alpha<0$ (resp.
$\alpha>0$), the azimuthal velocity is bigger (resp. smaller) than the
solid body rotation. Finally, it is possible to adjust $\Gamma_0$ to a
desired value by choosing the good curvature $\alpha$, in order to lower
the threshold for dynamo action.

\begin{table*}
\begin{center}
\begin{tabular}{|l|r|c|c|c|c|c|c|c|c|c|c|}
\hline
Impeller & $\alpha (^0)$ & $\langle P \rangle$ & $\langle T \rangle$ & $\Gamma_0=\frac{\langle P \rangle}{\langle T \rangle}$ & $\langle P \rangle.\langle T \rangle$ & $\langle H \rangle$ & $\cal V$ & $K_p$ & $MaDo$ & $R_m^c \; (w=0)$ & $R_m^c \; (w=0.4)$\\
\hline 
TM74$(-)$ & $-34$ & $0.15$ & $0.34$ & $0.46$ & $0.052$ & $0.43$ & $0.78$ & $0.073$ & $1.86$ & n.i. & n.i. \\
\hline
TM73$(-)$ & $-24$ & $0.16$ & $0.34$ & $0.48$ & $0.055$ & $0.41$ & $0.72$ & $0.073$ & $1.73$ & n.i. & n.i. \\
\hline 
TM71$(-)$ & $-14$ & $0.17$ & $0.33$ & $0.53$ & $0.057$ & $0.49$ & $0.73$ & $0.069$ & $1.79$ & n.i. & $197$ (o) \\
\hline 		 	
TM70 & $0$ & $0.18$ & $0.30$ & $0.60$ & $0.056$ & $0.47$ & $0.65$ & $0.061$ & $1.64$ & (1) &  (1) \\
\hline
TM71 & $+14$ & $0.19$ & $0.28$ & $\bf {0.69}$ & $0.053$ & $0.44$ & $0.64$ & $0.056$ & $1.66$ & ${\bf 179}$ & $51$ \\
\hline 	
TM73 & $+24$ & $0.20$ & $0.25$ & ${\bf 0.80}$ & $0.051$ & $0.44$ & $0.60$ & $0.053$ & $1.60$ & $180$ & ${\bf 43}$ \\
\hline 	
TM74 & +$34$ & $0.21$ & $0.24$ & $0.89$ & $0.050$ & $0.44$ & $0.58$ & $0.043$ & $1.65$ & $\infty$ & $44$ \\
\hline 		
\end{tabular}
\caption{Global hydrodynamical dimensionless quantities (see text for definitions) for the radius $R= 0.75$ impeller family, rotating anticlockwise ($+$), or clockwise ($-$) (see Fig. \ref{fig:turbine}). The last two columns present the thresholds for kinematic dynamo action with ($w=0.4$) and without ($w=0$) conducting layer. Optimal values appear in bold font. Most negative curvatures have not been investigated (n.i.) but TM71$(-)$, which presents oscillatory (o) dynamo instability for $R_m^c=197$ with $w=0.4$. (1): TM70 impeller ($\Gamma_0=0.60$) has a tricky behavior exchanging stability between steady modes, oscillatory modes and a singular mode mirror-symmetric with respect to the periodization introduced along $z$ and thus not physically relevant.}
\label{tab:zob}
\end{center}
\end{table*}

Without conducting shell, the optimal impeller is the TM71
($\Gamma_0=0.69$). But its threshold $R_m^c=179$ cannot be achieved in
the VKS2 experiment. So, we now have to find another way to reduce
$R_m^c$, the only relevant factor for the optimization.

\subsection{Optimal configuration to be tested in the VKS2 sodium experiment}
 
\begin{figure}[tbp]
\begin{center}
\includegraphics[clip,width=8cm]{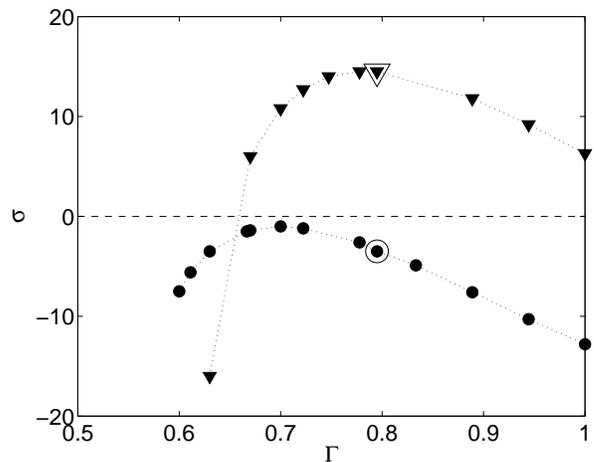}%
\end{center} 
\caption{Shift in the optimal value of $\Gamma$ when adding a conducting layer. Magnetic energy growth rate $\sigma \; vs. \; \Gamma$ for $w=0$ ($\bullet$) and $w=0.4$ ($\blacktriangledown$). Impeller TM73, $R_m=100$. Larger symbols mark the natural $\Gamma_0$ of the impeller.}
\label{fig:gammanumw04inset} 
\end{figure}

\begin{figure}[tbp]
\begin{center}
\includegraphics[clip,width=8cm]{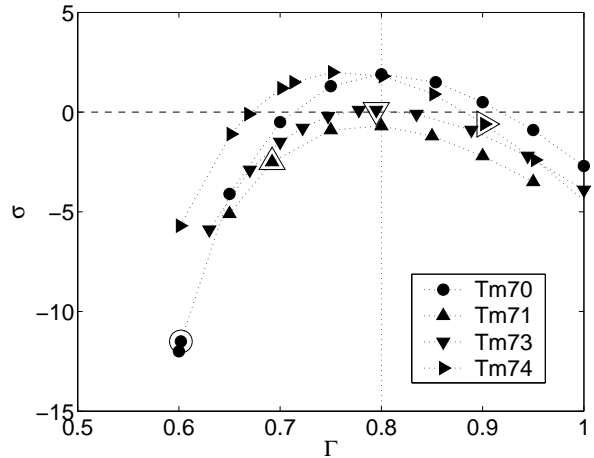}%
\end{center} 
\caption{Growth rate $\sigma$ of magnetic energy vs numerical ratio $\Gamma$. $R_m=43$, $w=0.4$ for $4$ different $R=0.75$ impellers: TM70 ($\bullet$), TM71 ($\blacktriangle$), TM73 ($\blacktriangledown$) and TM74 ($\blacktriangleright$). Larger symbols mark the natural $\Gamma_0$ of each impeller.} 
\label{fig:gammanumw04} 
\end{figure}

As in the Riga experiment \cite{stefani99,gailitis01}, and as in
numerical studies of various flows \cite{bullard77,kaiser99,avalos03},
we consider a stationary layer of fluid sodium surrounding the flow.
This significantly reduces the critical magnetic Reynolds number, but
also slightly shifts the optimal value for $\Gamma$. We have varied w
between $w=0$ and $w=1$; since the experimental VKS2 device is of fixed overall
size (diameter $0.6$ m), the flow volume decreases while increasing the
static layer thickness $w$. A compromise between this constraint and the
effects of increasing $w$ has been found to be $w=0.4$ and we mainly
present here results concerning this value of $w$. In
Fig.~\ref{fig:gammanumw04inset}, we compare the bell-shaped curves
obtained by numerical variation of the ratio $\Gamma$ for the same
impeller at the same $R_m$, in the case $w=0$, and $w=0.4$. The growth
rates are much higher for $w=0.4$, and the peak of the curve shifts from
$0.7$ to $0.8$. We have performed simulations for four different
velocity fields (Fig. \ref{fig:gammanumw04}), for $w=0.4$ at $R_m=43$: the result is very robust, the
four curves being very close.

\begin{figure}[tbp]
\begin{center}
\includegraphics[clip,width=8cm]{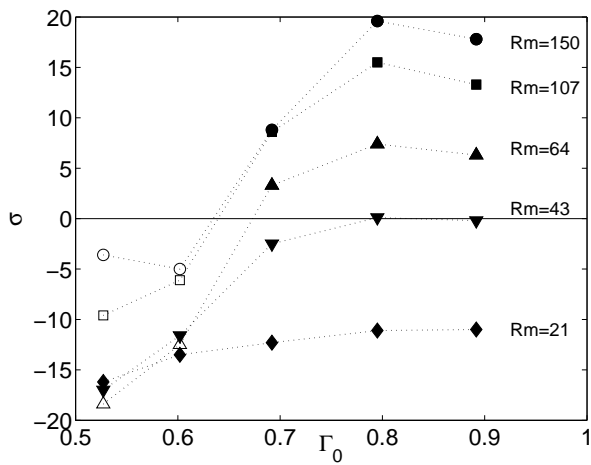}%
\end{center} 
\caption{Growth rate $\sigma$ vs natural ratio $\Gamma_0$ for five impellers at various $R_m$ and $w=0.4$. From left to right: TM71$(-)$ with $\Gamma_0=0.53$, TM70 ($\Gamma_0=0.60$), TM71 ($\Gamma_0=0.69$), TM73 ($\Gamma_0=0.80$), TM74 ($\Gamma_0=0.89$), see also table \ref{tab:zob}). Closed symbols: stationary modes. Open symbols: oscillating modes.}
\label{fig:gammareelw04} 
\end{figure}

In Fig. \ref{fig:gammareelw04}, we plot the growth rates $\sigma$ of the
magnetic energy simulated for four real mean velocity fields at various
$R_m$ and for $w=0.4$. The impeller TM73 was designed to create a mean
velocity field with $\Gamma_0=0.80$. It appears to be the best impeller,
with a critical magnetic Reynolds number of $R_m^c=43$. Its threshold is
divided by a factor $4$ when adding a layer of stationary conductor.
This configuration (TM73, $w=0.4$) will be the first one tested in the
VKS2 experiment. The VKS2 experiment will be able to reach the threshold
of kinematic dynamo action for the mean part of the flow. Meanwhile,
turbulence level will be high and could lead to shift or even
disappearance of the kinematic dynamo threshold. In 
Section~\ref{sec:layer}, we examine in details the effects of the
boundary conditions on TM73 kinematic dynamo.

\subsection{Role of flow helicity vs. Poloidal/Toroidal ratio} 

\begin{figure}[hbtp]
\begin{center}
\includegraphics[clip,width=8cm]{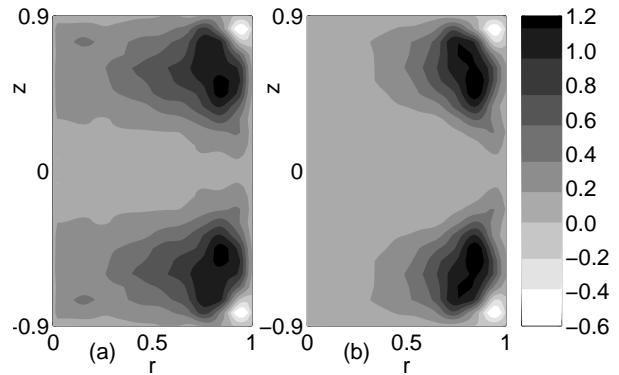}
\end{center} 
\caption{Isocontour of kinetic helicity $H={\bf v} . (\nabla \times {\bf v})$ for TM73 velocity field. (a): total helicity. (b): azimuthal contribution $v_{\theta} . (\nabla \times {\bf v})_{\theta}$ is dominant.}
\label{fig:helicity} 
\end{figure}

Most large scale dynamos known are based on helical flows
\cite{moffat78,parker55}. As a concrete example, while successfully
optimizing the Riga dynamo experiment, Stefani {\em et al.}
\cite{stefani99} noticed that the best flows were helicity maximizing.
The first point we focused on during our optimization process, {\em
i.e.}, the existence of an optimal value for $\Gamma$, leads us to
address the question of the links between $\Gamma$ and mean helicity
$\langle H \rangle$. In our case, for aspect ratio $H_c/R_c=1.8$ and
impellers of radius $R=0.75$, the mean helicity at a given rotation rate
$\langle H \rangle=\int {\bf v}.(\nabla \times {\bf v}) \; r dr dz$ does
not depend on the blade curvature (see Table \ref{tab:zob}). Observation
of Fig.~\ref{fig:helicity} also reveals that the dominant contribution
in the helicity scalar product is the product of the toroidal velocity
($v_{\theta} \propto \langle T \rangle$) by the poloidal recirculation
cells vorticity ($(\nabla \times {\bf v})_{\theta} \propto \langle P
\rangle$). We can therefore assume the scaling $\langle H \rangle
\propto \langle P \rangle \langle T \rangle$, which is consistent with
the fact that the product $\langle P \rangle \langle T \rangle$ and
$\langle H \rangle$ are both almost constant (Table \ref{tab:zob}).

\begin{figure}[htbp] 
\begin{center}
\includegraphics[clip,width=8cm]{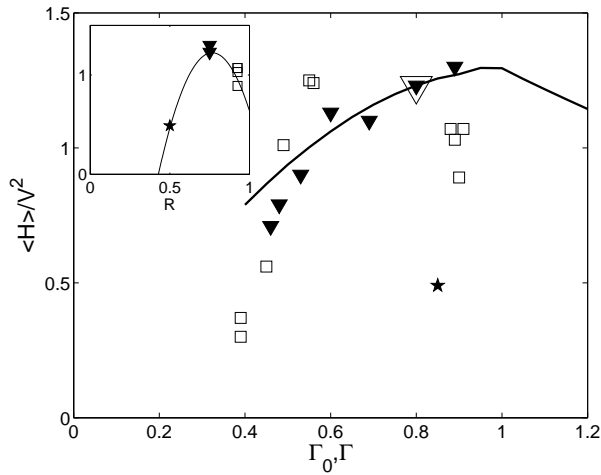} 
\end{center}
\caption{Mean helicity at a given $R_m$ ($\langle H \rangle/{\cal V}^2$)
{\em vs.} poloidal over toroidal ratio. The $R=0.75$ impeller
serie ($\blacktriangledown$) is plotted {\em vs.} $\Gamma_0$. The large open
symbol stands for TM73 at $\Gamma_0$ and the solid line stands for the
same quantity plotted {\em vs.} numerical variation of TM73 velocity field ($\Gamma$). 
We also plot $\langle H \rangle/{\cal V}^2$ {\em vs.} $\Gamma_0$ for the $R=0.5$ ($\star$) and $R=0.925$ ($\Box$) impellers.
The inset presents $\langle H \rangle/{\cal V}^2$ {\em
vs.} impeller radius $R$ for impellers of $0.8\alt\Gamma_0 \alt 0.9$. } 
\label{fig:heligam} 
\end{figure}

To compare the helicity content of different flows, we now consider the
mean helicity at a given $R_m$, $\langle H \rangle/ {\cal V}^2$, more
relevant for the dynamo problem. Figure \ref{fig:heligam} presents
$\langle H \rangle/{\cal V}^2$ {\em versus} $\Gamma_0$ for all $h=0.2$
impellers. The $R=0.75$ family reaches a maximum of order of $1$ for
$\Gamma_0 \simeq 0.9$. This tendency is confirmed by the solid curve
which stands for a numerical variation of $\Gamma$ for TM73 velocity
field and is maximum for $\Gamma=1$. Besides, even if $R=0.925$
impellers give reasonably high values of helicity near $\Gamma=0.5$,
there is an abrupt break in the tendency for high curvature: TM60 (see
Ref.~\cite{marie03}) exhibits large $\Gamma_0=0.9$ but less helicity
than TM74. Inset in Fig.~\ref{fig:heligam} highlights this optimum for
$\langle H \rangle/{\cal V}^2$ {\em versus} impeller radius $R$. This
confirms the impeller radius $R=0.75$ we have chosen during the optimization described above. 

As far as the optimal value toward dynamo action for the ratio $\Gamma$
(close to $0.7-0.8$, depending on $w$) is lower than $1$, the best velocity field is not
absolutely helicity-maximizing. In other words, the best dynamo flow
contains more toroidal velocity than the best helical flow. As shown by
Leprovost \cite{thesenico}, one can interpret the optimal $\Gamma$ as a
quantity that maximizes the product of mean helicity by a measure of
the $\omega$-effect, {\em i.e.}, the product $\langle H \rangle 
\langle T \rangle \sim \langle P \rangle \langle T \rangle ^2$.


\section{Impact of a conducting layer on the neutral mode and the energy
balance for the VKS2 optimized velocity field}
\label{sec:layer}

In this section, we deal with the mean velocity field produced between
two counterrotating TM73 impellers in a cylinder of aspect ratio
$\frac{H_c}{R_c}=1.8$, like the first experimental configuration chosen
for the VKS2 experiment. See Table \ref{tab:zob} for the characteristics
of this impeller, and Fig.~\ref{fig:ldv1} for a plot of the mean
velocity field. We detail the effects of adding a static layer of
conductor surrounding the flow and compare the neutral mode structures,
the magnetic energy and current density spatial repartition for this
kinematic dynamo.

\subsection{Neutral mode for $w=0$}

\begin{figure}[bp!]
\begin{center}
\includegraphics[clip,width=8cm]{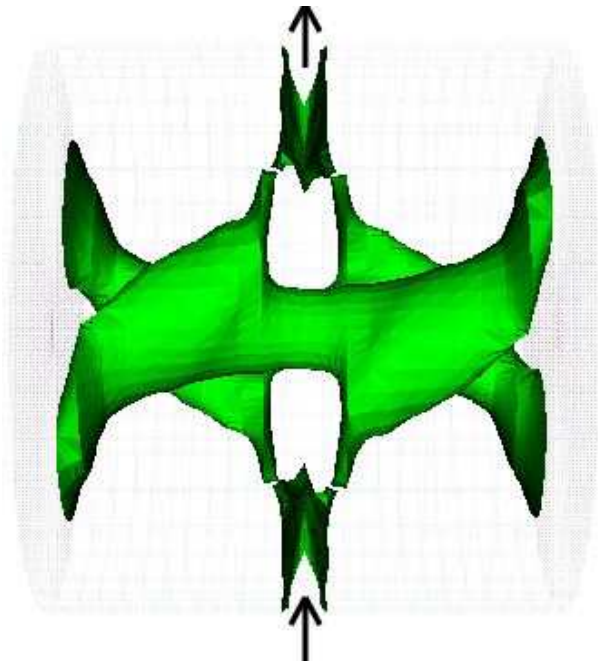}
\end{center}
\caption{Isodensity surface of magnetic energy ($50\%$ of the maximum) for the neutral mode without conducting layer ($w=0$). Cylinder axis is horizontal. Arrows stand for the external dipolar field source regions.}
\label{fig:isow0} 
\end{figure}

\begin{figure*}[!]
\begin{center}
\includegraphics[clip,width=17cm]{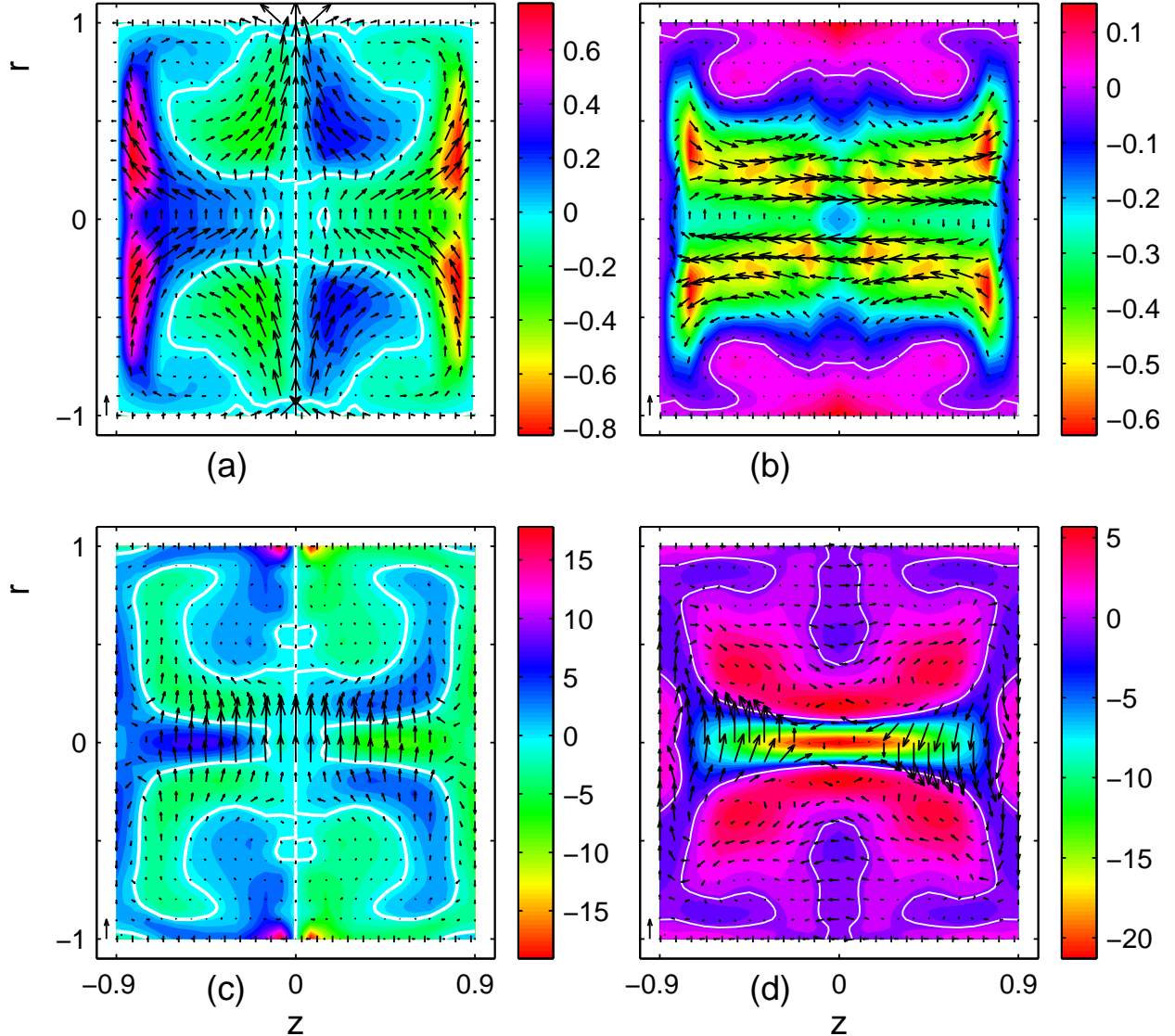}%
\end{center} 
\caption{Meridian sections of $\bf B$ and $\bf j$ fields for the neutral mode with $w=0$. $\bf B$ is divided by the total magnetic energy. Arrows correspond to components lying in the cut plane, and color code to the component transverse to the cut plane. A unit arrow is set into each figure lower left corner. (a): $\bf B$ field, $\theta=0$. (b) $\bf B$ field, $\theta=\frac{\pi}{2}$. (c): $\bf j$ field, $\theta=0$. (d): (c): $\bf j$ field, $\theta=\frac{\pi}{2}$.}
\label{fig:neutrew0} 
\end{figure*}

Without conducting layer, this flow exhibits dynamo action with a critical magnetic Reynolds number $R_m^c=180$. The neutral mode is stationary in time and has a $m=1$ azimuthal dependency. In Fig. \ref{fig:isow0}, we plot an isodensity surface of the magnetic energy ($50\%$ of the maximum) in the case $w=0$ at $R_m=R_m^c=180$. The field concentrates near the axis into two twisted banana-shaped regions of strong axial field. Near the interface between the flow and the outer insulating medium, there are two small sheets located on both sides of the plane $z=0$ where the magnetic field is almost transverse to the external boundary and dipolar. The topology of the neutral mode is very close to those obtained by Mari{\'e} {\em et al.} \cite{marie03} with different impellers, and to those obtained on analytical $s_2^0t_2^0-like$ flows in a cylindrical geometry with the previously described Galerkin analysis \cite{marie04}.

In Fig. \ref{fig:neutrew0} we present sections of the $\bf B$ and $\bf j$ fields, $\bf j=\bf \nabla \times \bf B$ being the dimensionless current density. The scale for $\bf B$ is chosen such as the magnetic energy integrated over the volume is unity. Since the azimuthal dependency is $m=1$, two cut planes are sufficient to describe the neutral mode. In the bulk where twisted-banana-shaped structures are identified, we note that the toroidal and poloidal parts of $\bf B$ are of the same order of magnitude and that $\bf B$ concentrates near the axis, where it experiences strong stretching due to the stagnation point in the velocity field. Around the center of the flow recirculation loops ($r \simeq 0.7$ and $z \simeq \pm 0.5$ see Fig.~\ref{fig:ldv1}) we note a low level of magnetic field: it is expelled from the vortices. Close to the outer boundary, we mainly observe a strong transverse dipolar field (Fig. \ref{fig:neutrew0} upper-left) correlated with two small loops of very strong current density $\bf j$ (Fig.~\ref{fig:neutrew0} lower-left). These current loops seem constrained by the boundary, and might dissipate great amount of energy by Joule effect (see discussion below).

\subsection{Effects of the conducting layer}

\begin{figure}[t]
\begin{center}
\includegraphics[clip,width=7cm]{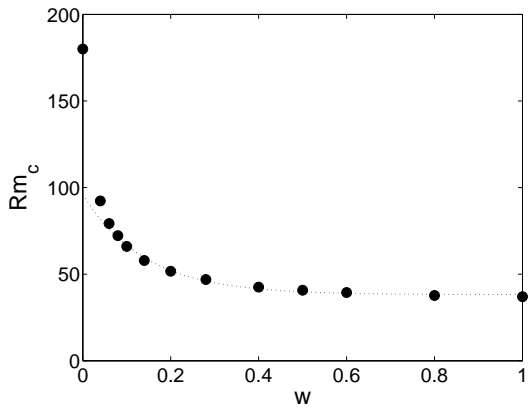}%
\end{center} 
\caption{Critical magnetic Reynolds number vs layer thickness $w$. TM73 velocity field. Fit: $R_m^c(w)=38+58\; exp(-\frac{w}{0.14})$ for $w \geq 0.08$.}
\label{fig:rmc} 
\end{figure}

\begin{figure}[tp!]
\begin{center}
\includegraphics[clip,width=7cm]{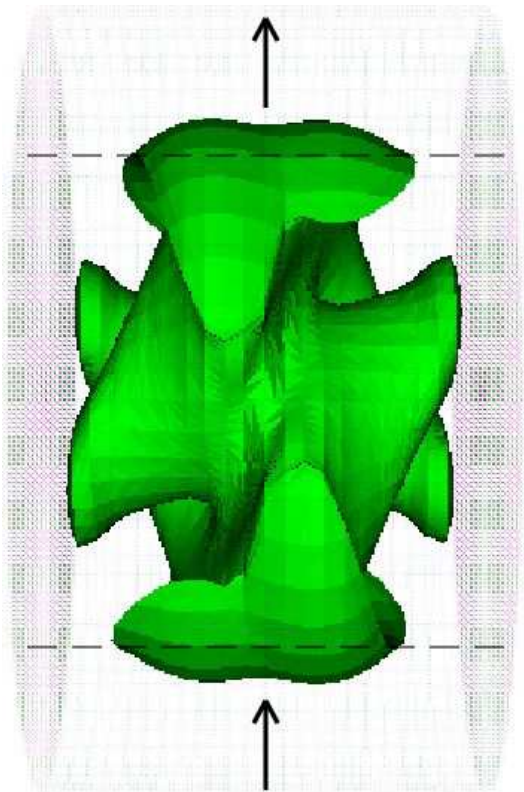}
\end{center}
\caption{Isodensity surface of magnetic energy ($50\%$ of the maximum) for the neutral mode with $w=0.6$.}
\label{fig:isow06} 
\end{figure}

\begin{figure*}[p!]
\begin{center}
\includegraphics[clip,width=17cm]{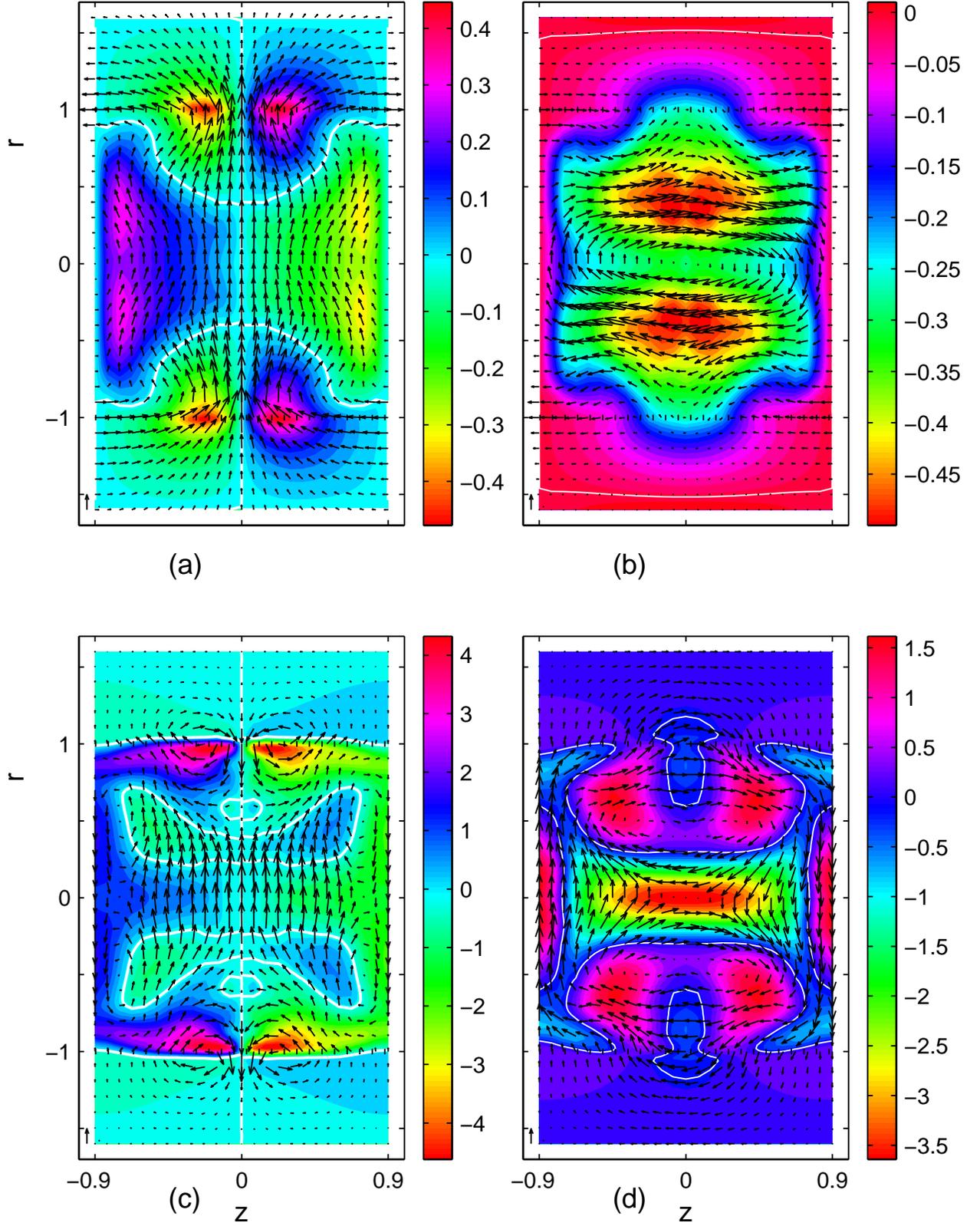}
\end{center} 
\caption{Meridian sections of $\bf B$ and $\bf j$ fields for the neutral mode with $w=0.6$. $\bf B$ is divided by the total magnetic energy. Arrows correspond to components lying in the cut plane, and color code to the component transverse to the cut plane. A unit arrow is set into each figure lower left corner. (a): $\bf B$ field, $\theta=0$. (b) $\bf B$ field, $\theta=\frac{\pi}{2}$. (c): $\bf j$ field, $\theta=0$. (d): (c): $\bf j$ field, $\theta=\frac{\pi}{2}$.}
\label{fig:neutrew06} 
\end{figure*}

As indicated in the first section, the main effect of adding a conducting layer is to strongly reduce the threshold. In Fig.~\ref{fig:rmc}, we plot the critical magnetic Reynolds number for increasing values of the layer thickness. The reduction is important: the threshold is already divided by $4$ for $w=0.4$  and the effects tends to saturate exponentially with a characteristic thickness $w=0.14$ (fit in Fig.~\ref{fig:rmc}), as observed for an $\alpha^2$-model of the Karlsruhe dynamo by Avalos {\em et al.} \cite{avalos03}. Adding the layer also modifies the spatial structure of the neutral mode: isodensity surface for $w=0.6$ is plotted in Fig. \ref{fig:isow06} with the corresponding sections of $\bf B$ and $\bf j$ fields in Fig. \ref{fig:neutrew06}. The two twisted bananas of axial field are still present in the core, but the sheets of magnetic energy near the $r=1$ boundary strongly develop. Instead of thin folded sheets on both sides of the equatorial plane, the structures unfold and grow in the axial and azimuthal directions to occupy a wider volume and extend on both sides of the flow/conducting-layer boundary $r=1$. This effect is spectacular and occurs even for low values of $w$. 

Small conducting layers are a challenge for numerical calculations: as far as the measured tangential velocity at the wall is not zero, adding a layer of conductor at rest gives rise to a strong velocity shear, which in practice needs at least 10 grid points to be represented. The maximal grid width used is $0.005$: the minimal non-zero $w$ is thus $w=0.05$. The exponential fit in Fig.~\ref{fig:rmc} is relevant for $w \agt 0.1$. We can wonder if the departure from exponential behavior is of numerical origin, or corresponds to a cross-over between different dynamo processes.

The analysis of $\bf B$ and $\bf j$ profiles in Fig.~\ref{fig:neutrew06} first reveals smoother $\bf B$-lines and much more homogeneous a repartition for the current density. The azimuthal current loops responsible for the transverse dipolar magnetic field now develop in a wider space (Fig.~\ref{fig:neutrew06} lower-left). Two poloidal current loops appear in this plane, closing in the conducting shell. These loops are responsible for the growth of the azimuthal magnetic field at $r=1$ (Fig.~\ref{fig:neutrew06} upper-left). Changes in the transverse plane ($\theta=\frac{\pi}{2}$) are less spectacular. As already stated in Refs.~\cite{kaiser99,avalos03}, the positive effect of adding a layer of stationary conductor may reside in the subtle balance between magnetic energy production and Ohmic dissipation. 


\subsection{Energy balance}
In order to better characterize which processes lead to dynamo action in a von K{\'a}rm{\'a}n flow, we will now look at the energy balance equation: let us first separate the whole space into three domains.

\begin{itemize} 
\item[$\bullet$] $\Omega_i: 0~<~r~<~1$ (inner flow domain) 
\item[$\bullet$] $\Omega_o: 1~<~r~<~1+w$ (outer stationary conducting layer)
\item[$\bullet$] $\Omega_\infty : r~>~1+w$ (external insulating medium) 
\end{itemize}

In any conducting domain $\Omega_\alpha$, we write the energy balance equation: 

$$\frac{\partial}{\partial t}\int_{\Omega_\alpha}{\bf{B}^2} = R_m \int_{\Omega_\alpha}{(\bf{j} \times \bf{B}).\bf{V}} - \int_{\Omega_\alpha}{\bf{j}^2} + \int_{\partial \Omega_\alpha}{(\bf{B}\times \bf{E}).\bf{n}}$$


The term in the left part of the equation is the temporal
variation of the magnetic energy $E_{mag}$. The first term in the right
part of the equation corresponds to the source term which writes as a
work of the Lorentz force. It exists only in $\Omega_i$ and is denoted
$W$. The second term is the Ohmic dissipation $D$, and the last term is
the Poynting vector flux $P$ which vanishes at infinite $r$.

We have checked our computations by reproducing the results of Kaiser and Tilgner \cite{kaiser99} on the Ponomarenko flow.

At the dynamo threshold, integration over the whole space gives $$0=W-D_o-D_i$$

\begin{figure}[hbp!]
\begin{center}
\includegraphics[clip,width=8cm]{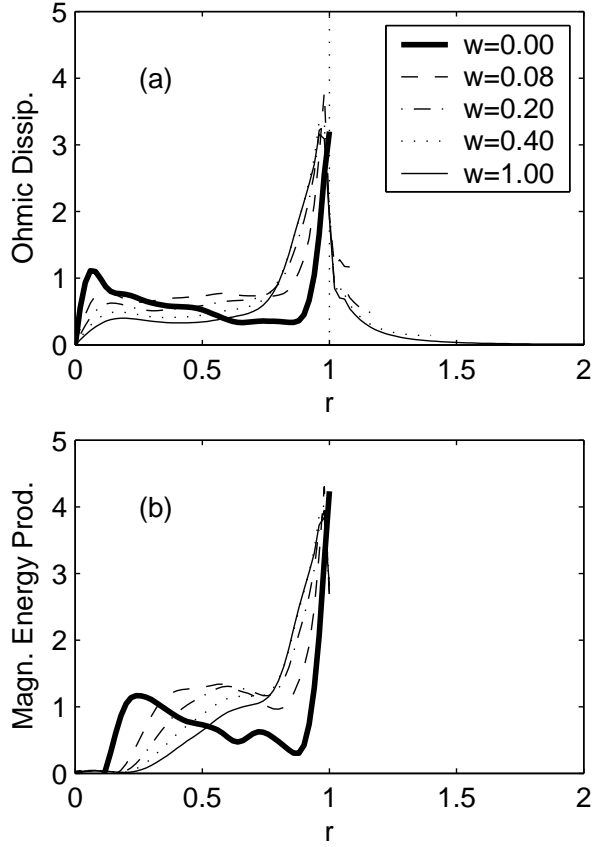}%
\end{center} 
\caption{(a): radial profile of Ohmic dissipation integrated other $\theta$ and $z$: $\int_{0}^{2\pi}\int_{-0.9}^{0.9} r\;{\bf{j}}^2(r)\;dz\;d\theta$ for increasing values of $w$.
(b): radial profile of magnetic energy production integrated other $\theta$ and $z$: $\int_{0}^{2\pi}\int_{-0.9}^{0.9} r\;((\bf {j} \times \bf {B}).\bf{V})(r)\;dz\;d\theta$ for increasing values of $w$.}
\label{fig:D} 
\end{figure}

In Fig. \ref{fig:D}, we plot the integrands of $W$ and $D$ at the
threshold for dynamo action, normalized by the total instantaneous
magnetic energy, as a function of radius $r$ for various $w$. For
$w=0$, both the production and dissipation mostly take place near the
wall between flow and the insulating medium ($r=1$), which could not
have been guessed from the cuts of $\bf j$ and $\bf B$ in figure
\ref{fig:neutrew0}: the $w=0$ curve in Fig. \ref{fig:D} has two bumps.
The first one at $r \simeq 0.1$ corresponds to the twisted bananas,
while the second is bigger and is localized near the flow boundary $r=1$. A lot of current
should be dissipated at the conductor-insulator interface due to the
``frustration'' of the transverse dipole. This can explain the huge
effect of adding a conducting layer at this interface: the ``strain
concentration'' is released when a conducting medium is added. So if we
increase $w$, the remaining current concentration at $r=1+w$ decreases
very rapidly to zero, which explains the saturation of the effect. In
the mean time, the curves collapse on a single smooth curve, both for
the dissipation and the production (solid black curves in Fig.
\ref{fig:D}). For greater values of $w$, the production density and the
dissipation in the core of the flow $r<0.2$ are smaller, whereas a peak
of production and dissipation is still visible at the flow-conducting
shell interface $r=1$. The conducting layer does not spread but
reinforces the localization of the dynamo process at this interface.
This can help us to understand the process which rises the dynamo in a
von K{\'a}rm{\'a}n type flow. 

\begin{figure}[tbp]
\begin{center}
\includegraphics[clip,width=8cm]{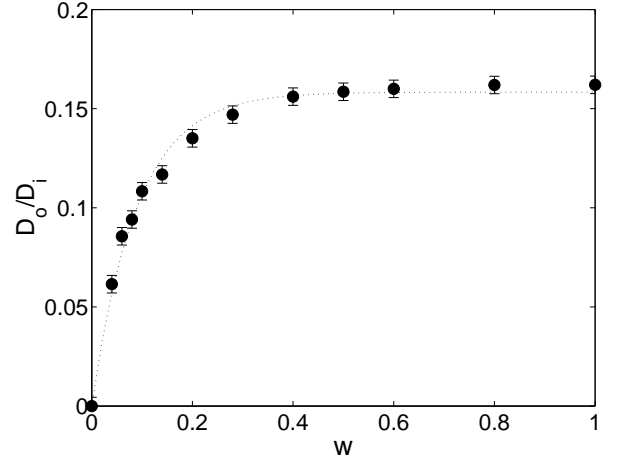}%
\end{center} 
\caption{Ratio of the integrated dissipation in the outer region and in the inner region $\frac{D_o}{D_i}$ vs $w$. Fit: $\frac{D_o}{D_i}(w)=0.16 \; (1 - exp(-\frac{w}{0.089}))$.}
\label{fig:DesurDi} 
\end{figure}

Let us now look at the repartition between the dissipation integrated
over the flow $D_i$ and the dissipation integrated over the conducting
shell $D_o$~(Fig. \ref{fig:DesurDi}). The ratio ${D_o}/{D_i}$ increases
monotonically with $w$ and then saturates to $0.16$. This ratio remains
small, which confirms the results of Avalos {\em et al.} \cite{avalos03}
for a stationary dynamo. We conclude that the conducting layer existence
---allowing currents to flow--- happens to be more important than
the relative amount of Joule energy dissipated in this layer.

\subsection{Neutral mode structure} 

From the numerical results presented above in this section, we consider
the following questions : Is it possible to identify typical structures
in the eigenmode of the von K\'arm\'an dynamo ? If yes, do these
structure play a role in the dynamo mechanism ? We have observed
magnetic structures in the shape of bananas and sheets (see
Figs.~\ref{fig:isow0} and \ref{fig:isow06}). In the center of the flow
volume, there is an hyperbolic stagnation point equivalent to
``$\alpha$-type'' stagnation points in ABC-flows (with equal
coefficients) \cite{childress95}. In the equatorial plane at the
boundary the merging of the poloidal cells remembers ``$\beta$-type''
stagnation points in ABC-flows. In such flows, the magnetic field is
organized into cigars along the $\alpha$-type stagnation points and
sheets on both sides of the $\beta$-type stagnation points
\cite{archontis03}: this is very similar to the structure of the neutral
mode we get for $w=0$ (Fig. \ref{fig:isow0}). We also performed magnetic
induction simulations with an imposed axial field for the poloidal part
of the flow alone. We obtain a strong axial stretching: the central
stagnation point could be responsible for the growth of the
bananas/cigars, which are twisted by the axial differential rotation
after. One should nevertheless not forget that real instantaneous flows
are highly turbulent, and that such peculiar stagnation points of the
mean flow are especially sensitive to fluctuations. 

\begin{figure}[tbp]
\begin{center}
\includegraphics[clip,width=8.5cm]{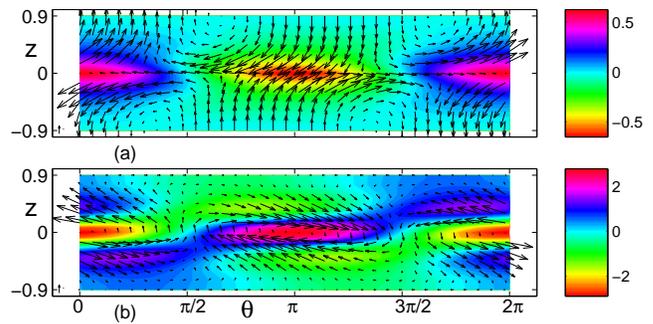}
\end{center} 
\caption{(a): (resp. (b)) $\bf B$ (resp. $\bf j$) field at $r=1$ for $w=0.6$. Color code corresponds to $B_r$ (resp. $j_r$) and arrows to $B_z$ and $B_{\theta}$ (resp. $j_z$ and $j_{\theta}$).}
\label{fig:developpee} 
\end{figure}

The presence of the conducting layer introduces new structures in the
neutral mode (see Figs.~\ref{fig:isow0}, \ref{fig:isow06} and
\ref{fig:neutrew0}, \ref{fig:neutrew06}). In order to complete our view
of the fields in the conducting layer, we plot them on the $r=1$
cylinder for $w=0.6$ (Fig.~\ref{fig:developpee}). 
As for $w=0$, the dipolar main part of the magnetic field gets radially into the flow
volume at $\theta=\pi$ and exits at $\theta=0$ (Fig.
\ref{fig:developpee} up). However, looking around $z=0$, we observe that a part
of this magnetic flux is azimuthally diverted in the conducting shell along the flow
boundary. This effect does not exist without conducting shell: the outer
part of the dipole is anchored in the stationary conducting layer.

Another specific feature is the anti-colinearity of the current density
$\bf j$ with $\bf B$ at ($z=0;\theta=0$,$\pi;r=1$), which could remind an
``$\alpha$''-effect. However, while the radial magnetic field is clearly
due to a current loop (arrows in the center of Fig.~\ref{fig:developpee} down), $j_r$ is not linked
to a $\bf B$-loop (Fig.~\ref{fig:developpee} up), which is not obvious from Fig.~\ref{fig:neutrew06}.
Thus, the anti-colinearity is restricted to single points
($z=0;\theta=0,\pi ; r=1$). We have checked this, computing the angle between $\bf j$ and $\bf B$: the isocontours of this angle are very complex and the peculiar values corresponding to colinearity or anti-colinearity are indeed restricted to single points. 

\subsection{Dynamo threshold reduction factor}

We have shown that the threshold for dynamo action is divided by four
when adding a conducting layer of thickness $w=0.4$. This effect is very
strong. Following Avalos and Plunian \cite{avalos03}, let us compare the
threshold reduction factor $\Lambda=1-\frac{R_m^c(w)}{R_m^c(w=0)}$ for
various kinematic dynamos. The threshold reduction for TM73-flow
($\Lambda=0.78$) is much higher than for the Karlsruhe ($\Lambda=0.11$)
and Riga ($\Lambda=0.56$) dynamos. Reduction rate can also be radically
different between model flows: the $\alpha^2$-model
for Karlsruhe dynamo gives a low-$R_m^c$-dynamo for $w=0$ and benefits
very few of a finite $w$ ($\Lambda=0.11$), while the Ponomarenko flow does not lead to dynamo
action without conducting layer ($\Lambda=1$). The reduction factors
considered above are maximal values obtained either for high $w$ in
stationary dynamos or for the optimal $w$ in oscillatory dynamos \cite{kaiser99,avalos03}.

In order to understand why $\Lambda$ is so high for our TM73-flow, we
propose to compare our experimental flow with an optimal analytical
model-flow proposed by Mari{\'e}, Normand and Daviaud~\cite{marie04} in
the same geometry. The Galerkin method used by these authors does not
allow to study the effect of a conducting layer. We thus perform
kinematic dynamo simulations with our usual approach, and then study the
effects of adding a conducting layer on the following velocity field for
$\epsilon=0.7259$ corresponding to $\Gamma=0.8$
\cite{theselouis,marie04}: 

\begin{eqnarray}
v_r & = & -\frac{\pi}{2} r(1-r)^2(1+2r) \cos(\pi z) \nonumber \\ 
v_{\theta} & = & 4 \epsilon r(1-r) \sin(\pi z/2)  \nonumber \\
v_z & = &(1-r)(1+r-5r^2) \sin(\pi z) \label{eq:mnd}
\end{eqnarray}

\noindent The kinematic dynamo threshold is found at $R_m^c=58$ for
$w=0$, in good agreement with the galerkin analysis. With a $w=1$
conducting layer, we get a low $\Lambda=0.26$ reduction rate, {\em i.e.}
$R_m^c=43$, close to the TM73 threshold for $w=1$: $R_m^c=37$. The
threshold reduction is also found to show an exponential behavior with
$w$, of characteristic thickness $0.20$, as in Fig.~\ref{fig:rmc}.

Let us describe the model flow features represented in Fig.~\ref{fig:radialV} (bottom).
The velocity is very smooth at the cylindrical boundary: the toroidal
velocity is maximum at $r=0.5$ and slowly decreases to zero at $r=1$.
The poloidal recirculation loops are centered at $r_p=0.56$ and the
axial velocity also decreases slowly to zero at the cylindrical
boundary. Thus, the mass conservation requires the axial velocity to be much
higher in the central disk ($0<r<r_p$) than outside. 
These constraints make analytical models somewhat different from
experimental mean flows (Fig.~\ref{fig:radialV}). In particular, high
kinetic Reynolds numbers forbid smooth velocity decrease near
boundaries. This explains why experimental flows do not lead to low
thresholds unless a conducting layer is added.

We now consider the effect of a conducting shell on the model flow
eigenmode structure. First note that without conducting shell, the model
neutral mode structure is already very similar to that of TM73 with
conducting shell: the transverse dipole is not confined into thin sheets
but develops into wider regions connected to bananas of axial field in
the center. Adding the conducting layer mainly lets the neutral mode
structure unchanged and thus quantitatively reduces its impact compared
to the experimental case.

Finally, from the very numerous simulations of experimental and model
von K\'arm\'an flows performed, we conclude that the adjunction of a
static conducting layer to experimental flows makes the eigenmode
geometry closer to optimal model eigenmodes, and critical $R_m^c$ get
closer to moderate values (typically $50$). It may thus be conjectured
that the puzzling sensitivity of dynamo threshold to flow geometry is
lowered when a static layer is present. We conclude this feature renders
the dynamo more robust to flow topology details. This could also act
favorably in the nonlinear regime.

\section{Conjectures about dynamo mechanisms}
\label{sec:conjec}

In this paragraph, we intend to relate the results of the optimization
process to some more elementary mechanisms. As emphasized in the
Introduction, there is no sufficient condition for dynamo action and
although numerical examples of dynamo flows are numerous, little is
known about the effective parameters leading to an efficient energy
conversion process. For example, the classical $\alpha$ and axial
$\omega$ mechanisms have been proposed to be the main ingredients of the
von K\'arm\'an dynamo \cite{marie02}. Our starting point is the
observation that dynamo action results from a constructive coupling
between magnetic fields components due to velocity gradients, which, in
the present axisymmetric case, reduce to derivatives with respect to $r$
(radial gradients) and to $z$ (axial gradients). The gradients of
azimuthal velocity generate a toroidal field from a poloidal one
($\omega$-effect \cite{moffat78}), while regeneration of the poloidal
field is generally described as resulting from an helicity effect
(denoted $\alpha$-effect if scale separation is present
\cite{krause80}). How do these general considerations apply to the
present flow ? As in the Sun, which shows both a polar-equatorial
differential rotation and a tachocline transition, our experimental flow
fields present azimuthal velocity shear in axial and radial directions
(see Fig.~\ref{fig:ldv1}). So, we will consider below the role of both
axial and radial $\omega$-effect. 

We will discuss these mechanisms and then suggest that, for a flow surrounded by a
static conducting layer, the dynamo mechanism is based
on the presence of a strong velocity shear (at the boundary layer $r =
1$) which lies in this case in the bulk of the overall electrically
conducting domain.

\subsection{Axial $\omega$-effect}

Induction simulations performed with the toroidal part of the velocity
show an axial $\omega$-effect which converts an imposed axial field into
toroidal field through $\partial v_{\theta} / \partial z$. Such a
$R_m$-linear effect has been evidenced in VKS1 experiment
\cite{bourgoin02}. This effect concentrates around the equatorial shear
layer ($z=0$) as visible in Fig.~\ref{fig:ldv1}. Thus, we can think that
the axial $\omega$-effect is involved in the dynamo process: for dynamo
action to take place, there is a need for another process to convert
toroidal magnetic field into poloidal field.

\subsection{$\alpha$-effect, helicity effect}

$R_m$-non-linear conversion from transverse to axial magnetic field has
also been reported in VKS1 experiment \cite{petrelis02}. This effect is
not the usual scale-separation $\alpha$-effect \cite{krause80} and has
been interpreted as an effect of the global helicity as reported by
Parker \cite{parker55} (in the following, it will be denoted
``$\alpha$''-effect). We believe it to take place in the high kinetic helicity
regions of the flow (see Fig.~\ref{fig:helicity}).

\subsection{Is an ``$\alpha$''$\omega$ mechanism relevant ?}

Bourgoin {\em et al.} \cite{bourgoin04} performed a study of induction
mechanisms in von K{\'a}rm{\'a}n-type flows, using a quasi-static
iterative approach. They show that ``$\alpha$''$\omega$ dynamo action,
seen as a three-step loop-back inductive mechanism, is possible, but
very difficult to obtain, fields being widely expelled by the vortices.
The authors highlight the fact that the coupling between the axial
$\omega$-effect and the ``$\alpha$''-effect is very inefficient for our
velocity fields, because of the spatial separation of these two
induction effects. Our observations of the velocity and helicity fields
confirm this separation.

The authors also discovered an induction effect --- the BC-effect --- related
to the magnetic diffusivity discontinuity at the insulating boundary
that could be invoked in the dynamo mechanism. This BC-effect, illustrated on our
TM73-velocity field (Fig.~14 in Ref.~\cite{bourgoin04}), is enhanced in
the case of strong velocity and vorticity gradients at the boundaries,
characteristic of high Reynolds number flows. So, we are convinced that
for experimental flow fields at $w=0$, the BC-effect helps the dynamo.
This is coherent with our observations of high tangential current
density near the boundaries and high magnetic energy production at $r=1$
even for $w=0$ (Fig.~\ref{fig:D}). Such a current sheet formation and
BC-effect was reported by Bullard and Gubbins \cite{bullard77}.

When a large layer of sodium at rest is added, the
BC-effect vanishes because the conductivity discontinuity occurs at
$r=1+w$ while the currents still concentrate at the flow boundary $r=1$.
However, with a conducting layer, we have presented many features
favoring the dynamo: in the next paragraph, we propose a possible
origin for this conducting-layer effect. 

\subsection{Radial $\omega$-effect, boundary layers and static shell}

With a layer of steady conducting material surrounding the flow, we
note the occurrence of two major phenomena:
\begin{itemize}

\item the possibility for currents to flow freely in
this shell (Fig. \ref{fig:D}), 
\item the presence of a very strong velocity shear localized
at the boundary layer which now lies in the bulk of the electrically
conducting domain. 

\end{itemize}
 
Let us again consider the shape of the velocity shear. Any realistic
(with real hydrodynamical boundary conditions) von K\'arm\'an flow
obviously presents negative gradients of azimuthal velocity $\partial
v_{\theta}/ \partial r$ between the region of maximal velocity and the
flow boundary. This region can be divided into two parts: a smooth
decrease in the bulk ($R \alt r \alt 1$) and a sharp gradient in the
boundary layer at $r=1$ (Fig.~\ref{fig:radialV}).

These gradients are responsible for a radial
$\omega$-effect, producing $B_{\theta}$ with $B_r$, in both insulating
and conducting cases. However, without conducting layer, only the smooth
part of the gradient which lies in the bulk will be efficient for dynamo
action. Indeed, owing to the huge value of the kinetic Reynolds number
and the very small value of the magnetic Prandtl number, the sharp
boundary layer gradient is confined in a tiny domain, much smaller than the
magnetic variation scale. No significant electrical currents can
flow in it and we did not resolve this boundary layer with the
numerical code: it is totally neglected by our approach.

The role of both types of gradients is illustrated by the observation
(Fig.~\ref{fig:radialV}, left) of impellers of large radius ($R=0.925$).
For such impellers there is almost no departure from solid body rotation
profiles in the flow region and these impellers lead to dynamo action
only with conducting shell \cite{marie03}, {\em i.e.}, due to the sharp
gradient. On the other hand, our
$R=0.75$ selected impellers present a stronger bulk-gradient and
achieve dynamo in both cases.

Actually, the way we numerically modelized the von K\'arm\'an flow
surrounded by a static conducting layer ---considering an equivalent fluid
system in which the boundary layer appears as a simple velocity jump in
its bulk--- is coherent with the problem to solve. The velocity jump, just
as any strong shear, is a possible efficient source for the radial
$\omega$-effect.

\subsection{A shear and shell dynamo ?}

We pointed out above that the regions of maximal helicity (the
``$\alpha$''-effect sources, see Fig.~\ref{fig:helicity}) are close to
those of radial shear where radial $\omega$-effect source term takes
place. Dynamo mechanism could thus be the result of this interaction.
So, in the absence of a static shell, one can suppose that the dynamo arises from
the coupling of ``$\alpha$''-effect, $\omega$-effect and the BC-effect
\cite{bourgoin04}. With a static conducting layer, as explained above,
the radial $\omega$-effect is especially strong: the radial
dipole, anchored in the conducting layer and azimuthally stretched by the
toroidal flow (see Fig.~\ref{fig:developpee}) is a strong source of
azimuthal field. This effect coupled with the ``$\alpha$''-effect  could
be at the origin of the dynamo.

For small conducting layer thickness $w$, one could expect a cross-over
between these two mechanisms. In fact, it appears that the decrease of
$R_m^c$ (Fig.~\ref{fig:rmc}) with the conducting shell thickness $w$ is
very fast between $w=0$ and $w=0.08$ and is well fitted for greater $w$ by an
exponential, as in Ref.~\cite{avalos03}. We can also note that for
typical $R_m=50$, the dimensionless magnetic diffusion length
$R_m^{-1/2}$ is equal to $0.14$. This value corresponds to the
characteristic length of the $R_m^c$ decrease (Fig.~\ref{fig:rmc}) and is also close to the
cross-over thickness and characteristic lengths of the Ohmic dissipation profiles (Figs.~\ref{fig:D}, top and \ref{fig:DesurDi}).

We propose to call the mechanism described above a ``shear and shell''
dynamo. This interpretation could also apply to the
Ponomarenko screw-flow dynamo which also merely relies on the presence
of an external conducting medium.

\section{Conclusion}

We have selected a configuration for the mean flow feasible in the VKS2 liquid
sodium experiment. This mean flow leads to kinematic dynamo action for a
critical magnetic Reynolds number below the maximum achievable $R_m$. We
have performed a study of the relations between kinematic dynamo action,
mean flow features and boundary conditions in a von K\'arm\'an-type
flow. 

The first concluding remark is that while the dynamo without static
conducting shell strongly depends on the bulk flow details, adding a
stationary layer makes the dynamo threshold more robust. The study of
induction mechanisms in 3D cellular von K\'arm\'an type flows performed
by Bourgoin {\em et al.} \cite{bourgoin04} suggests that this
sensitivity comes from the spatial separation of the different induction
mechanisms involved in the dynamo process: the loop-back between these
effects cannot overcome the expulsion of magnetic flux by eddies if the
coupling is not sufficient. Secondly, the role of the static layer is
generally presented as a possibility for currents to flow more freely.
But, instead of a spreading of the currents, the localization at the
boundary of both magnetic energy production and dissipation
(Fig.~\ref{fig:D}) appears strongly reinforced. Actually, strong shears
in the bulk of the electrically conducting domain imposed by material
boundaries are the dominating sources of dynamo action. They result in a
better coupling between the inductive mechanisms. We also notice that
there seems to be a general value for the minimal dynamo threshold
(typically $50$) in our class of flows, for both best analytical flows
and experimental flows with static conducting layer.

Although the lowering of the critical magnetic Reynolds number due to an
external static envelope seems to confirm previous analogous results
\cite{thesepetrelis,kaiser99,avalos03}, it must not be considered as
the standard and general answer. In fact, in collaboration with Frank
Stefani and Mingtian Xu from the Dresden MHD group, we are presently
examining how such layers, when situated at both flat ends, {\em i.e.},
besides the propellers, may lead to some increase of the critical
magnetic Reynolds number. This option should clearly be avoided to
optimize fluid dynamos similar to VKS2 configuration. However, a
specific study of this later effect may help to better understand how
dynamo action, which is a global result, relies also on the mutual effects of
separated spatial domains with different induction properties. 

\begin{acknowledgments} We would like to thank the other members of the
VKS team, M. Bourgoin, S. Fauve, L. Mari{\'e}, P. Odier, F.
P{\'e}tr{\'e}lis, J.-F. Pinton and R. Volk, as well as B.~Dubrulle,
N.~Leprovost, C.~Normand, F.~Plunian and F.~Stefani for fruitful
discussions. We are indebted to V.~Padilla and C.~Gasquet for technical
assistance. We thank the GDR dynamo for support. \end{acknowledgments}

\end{document}